\documentclass{aa}  
\bibpunct{(}{)}{;}{a}{}{,} 
\usepackage{graphicx}
\usepackage{xcolor,lastpage}
\usepackage[colorlinks=true,linkcolor=blue,citecolor=blue]{hyperref}
\usepackage{txfonts,siunitx}
\begin{document}

   \title{Discovery of a multiphase \ion{O}{VI} and \ion{O}{VII} absorber in the circumgalactic/intergalactic transition region}
   

\author{Jussi Ahoranta\inst{1}
	\and Alexis Finoguenov\inst{1}
	\and Massimiliano Bonamente\inst{2,3}
	\and Evan Tilton\inst{4} 
	\and Nastasha Wijers\inst{5}
    \and Sowgat Muzahid \inst{6,7}
	\and Joop Schaye\inst{5}
}

\institute{Department of Physics, University of Helsinki, P.O. Box 64, FI-00014, Finland
	\and Department of Physics, University of Alabama in Huntsville, Huntsville, AL, USA
	\and NASA National Space Science and Technology Center, Huntsville, AL, USA
	\and Deparment of Physics \& Astronomy, Regis University, Denver, CO 80221, USA
	\and Leiden Observatory, Leiden University, P.O. Box 9513, NL-2300 RA Leiden, the Netherlands
	\and IUCAA, Post Bag 04, Ganeshkhind, Pune-411007, India
    \and Leibniz-Institut f\"{u}r Astrophysik Potsdam (AIP), An der Sternwarte 16, D-14482 Potsdam, Germany
}

   \date{Received; accepted}

 
  \abstract
   {} 
   {The observational constraints on the baryon content of the warm-hot intergalactic medium (WHIM) rely almost entirely on far ultraviolet (FUV) measurements. However, cosmological, hydrodynamical simulations predict strong correlations between the spatial distributions of FUV and X--ray absorbing WHIM. We investigate this prediction 
   by analyzing XMM-Newton X--ray counterparts of FUV-detected intergalactic \ion{O}{VI} absorbers known from FUSE and HST/STIS data, thereby aiming to gain
   understanding on the properties of the hot component of FUV absorbers and to compare this information to the predictions of simulations. 
   }
   {
   We study the X--ray absorption at the redshift of the only significantly detected \ion{O}{vi} absorber 
   in the Ton~S~180 sightline's FUV spectrum, found at $z_\mathrm{OVI}=0.04579\pm0.00001$.
   We characterize the spectral properties of the \ion{O}{VI}-\ion{O}{VIII} absorbers and explore 
   the ionization processes behind the measured absorption. The observational results are compared to the predicted warm-hot gas properties in the \texttt{EAGLE} simulation to infer the physical conditions of the absorber.
   } 
   {We detect both \ion{O}{vi} and \ion{O}{vii} absorption at a $5\sigma$ confidence level,  whereas \ion{O}{viii} absorption is not significantly detected. Collisional ionization equilibrium (CIE) modeling constrains the X--ray absorbing gas temperature to log$\,T_\mathrm{CIE}~\mathrm{(K)}=6.22\pm0.05$ with a total hydrogen column density $N_\mathrm{H}=5.8_{-2.2}^{+3.0}\times Z_{\sun}/Z_\mathrm{abs}\times10^{19}$~cm$^{-2}$. 
   This model predicts an \ion{O}{vi} column density consistent with that measured in the FUV, but our limits on the \ion{O}{vi} line width 
   indicate $>90$~\% likelihood that the FUV-detected \ion{O}{vi} arises from a different, cooler phase.
   We find that the observed absorber lies about a factor of two further away
   from the detected galaxies than is the case for similar systems in \texttt{EAGLE}}
   {The analysis suggests that the detected \ion{O}{VI} and \ion{O}{VII} trace two different -- warm and hot -- gas phases 
   of the absorbing structure at $z\approx0.046$, of which the hot component is likely in collisional ionization equilibrium. 
   As the baryon content information of the studied absorber is primarily imprinted in the X--ray band, understanding the abundance of similar systems helps to define the landscape for WHIM searches with future X--ray telescopes.
   Our results highlight the 
   crucial role of line widths for the interpretation and detectability of WHIM absorbers. 
   }
 

\maketitle


\section{Introduction}

About half of the baryonic matter in the low-$z$ Universe is expected to exist in the form of tenuous, shock--heated, highly--ionized gas that accumulated in dark matter dominated large--scale structures, commonly referred to as the warm--hot intergalactic medium (WHIM) \citep[e.g.,][]{cen99, dave99,shull12,nicastro16}. Unlike the warm ($T\sim10^{5-6}$~K) WHIM phase that is more readily detected in the far ultraviolet (FUV) \citep[e.g,][]{2011ApJ...743..180S,2015MNRAS.446.2444H,2016MNRAS.458..733P}, the hot phase ($T\sim10^{6-7}$~K) remains practically undetected, presumably due to the lower sensitivity of the available
X--ray instruments to the weak absorption lines expected to trace this phase. Although a number of observations of 
hot WHIM systems have been published \citep[e.g.,][among others]{fang13,max16,nicastro18,kovacs19,ahoranta19}, the detection significances are typically marginal.

The most important ions for detecting spectral signals of the hot ($T\sim10^6$~K) WHIM phase are the highly--ionized oxygen species \ion{O}{VII} and \ion{O}{VIII}. Since the rest wavelengths of the strongest \ion{O}{VII} and \ion{O}{VIII} lines are located in the soft X--ray band, the properties of hot WHIM absorbers are best studied using high energy--resolution X--ray instruments, such as the XMM-Newton Reflection Grating Spectrometer (RGS) and the Chandra Low Energy Transmission Grating (LETG) spectrometer. However, the \texttt{EAGLE} cosmological, hydrodynamical simulation \citep[][]{schaye15,crain15,mcalpine16} predicts that
most of the signals from hot WHIM absorbers fall below the current X--ray detection threshold ($N_\mathrm{ion}\gtrsim10^{15}\,\mathrm{cm}^{-2}$) and that the abundance of the absorbers begins to drop steeply at the $N_\mathrm{ion}\sim10^{16}\,\mathrm{cm}^{-2}$ level \citep[e.g.,][]{wijers19}. Furthermore, the \texttt{EAGLE} simulation indicates that a large fraction of high column density \ion{O}{VII-VIII} absorbers within the cosmic web are associated with the gaseous halos surrounding galaxies (i.e., the circumgalactic medium, CGM), whereas the X--ray detectable absorbers of the intergalactic medium (IGM) are expected to become much rarer with increasing distance from the galactic halos \citep[][]{2019ApJ...884L..31J}.

The rarity of strong X--ray WHIM absorbers and the low expected detection significances of the associated ionic lines reduce the likelihood of identifying X--ray absorbing WHIM systems through a pure blind search. 
Fortunately, the \texttt{EAGLE} simulation also predicts correlations between the observables of the warm WHIM phase and the hot, X--ray absorbing phase (e.g., between $N_\mathrm{OVI}$ and $N_\mathrm{OVII}$; \citealt{2003ApJ...594...42C,wijers19}), which can be used to limit hot WHIM searches to the most likely locations.

In this paper, we leverage the prediction of correlated FUV and X--ray absorption, and we study the only significantly detected \ion{O}{VI} absorber ($z_\mathrm{OVI}\approx0.0458$) in the sightline toward Ton~S~180 \citep[][]{danforth06,tilton12}, a type 1 narrow-line Seyfert galaxy (NLS1) located at $z=0.0617$. The spectral properties of Ton~S~180 have been studied in multiple wavelength bands, 
revealing no evidence of $\gtrsim$\SI{e3}{\km\per\s} intrinsic outflows \citep[see e.g.,][]{turner02,gofford13,2013MNRAS.431.2885M}. The closest detected galaxy to the examined \ion{O}{VI} absorber is a late type galaxy with $L=0.4\,L^*$, located at $z=0.0456$ with an impact parameter $\rho\approx290$~pkpc to the sightline. It is the only $L>0.1L^*$ galaxy within $\rho<300$~kpc near the redshift of the detected \ion{O}{vi} absorber \citep{prochaska11},
and as discussed in this work,
unlikely to produce X--ray detectable column densities by itself. These characteristics of the sightline towards Ton~S~180 imply that any detected X--ray absorption lines would be likely associated with intervening WHIM.

This work is a part of a program focused on placing limits on the X--ray band metal absorption at the redshifts of the established \ion{O}{VI} absorbers. The full program sample includes all sightlines which have been observed at high spectral resolution both in FUV and X--ray bands. We report the results from the Ton~S~180 sightline separately from the rest of the sample, as we have found it to contain an outstandingly strong \ion{O}{VII} signal with an associated \ion{O}{VI} absorber, thus enabling more detailed analysis methods than would typically be applicable.

\begin{table}[ht]
\begin{center}
\label{table:observations}
\caption{Summary of XMM-Newton observations}
\begin{tabular*}{0.85\columnwidth}{p{0.3\columnwidth}p{0.2\columnwidth}p{0.2\columnwidth}}
\hline
\hline
 &  \multicolumn{2}{c}{Clean time (ks)} \\  
OID & RGS1 & RGS2 \\
\hline
0110890401 & 30.1 & 29.1  \\
0110890701 & 18.0 & 17.3 \\
0764170101 & 124.7 & 124.3 \\
0790990101 & 30.7 & 30.5 \\
\hline
Total & 203.5 & 201.2 \\
\hline
\end{tabular*}
\end{center}
\tablefoot{Observation identifiers and solar--flare--removed clean exposure times (average per channel) of the reduced RGS data.}
\end{table}

\section{Observations and data preparation}

\subsection{X-ray data}\label{x-rayre}

Ton~S~180 has been observed with XMM-\emph{Newton} RGS on four occasions, 
totaling $\sim 200$~ks exposure times for both instruments (Table \ref{table:observations}). 
In addition, there is one Chandra LETG ACIS-S observation (exp. time $\approx77$~ks). Despite the relatively high X--ray luminosity of Ton~S~180 during the Chandra exposure ($F_{0.5-2\,\mathrm{keV}}\approx1.2\times10^{-11}$~erg$\,$cm$^{-2}\,$s$^{-1}$), the low effective area of LETG ACIS-S at $\lambda\sim22$~\AA\ ($A_{eff}^\mathrm{ACIS-S}<10$~cm$^{2}$ compared to $A_{eff}^\mathrm{RGS1}\approx35$~cm$^{2}$) results in only a few first--order counts per channel at the wavelengths around the predicted \ion{O}{VII} He$\alpha$ location. Therefore we did not considered Chandra data in this work.

The RGS first-- and second--order data were processed using 
SAS v.18.0.0 with the latest calibration file libraries (up to XMM-CCF-REL-372). 
The data were reduced with the \texttt{rgsproc} pipeline adopting the default settings with three exceptions: 
(1) cool pixels, as identified by the \texttt{rgsbadpix} task, were discarded to exclude possible
spectral artifacts due to unaccounted pointing shifts within the observations; 
(2) aspect drift--based grouping of the data was performed with twice the accuracy 
of the default setting, and 3) the grating line--spread functions were calculated with the full convolution space.    
In addition, we examined the light-curves extracted from the RGS CCD9 chips to identify time periods of
Solar flares/radiation belt transits in the observational data. The good time interval files were then generated with use of the \texttt{tabgtigen} "rate" option and employed in the data filtering to exclude the 
time periods of elevated high energy particle fluxes in the detector plane.

The RGS spectra were converted into the \texttt{SPEX} \citep{kaastra1996_spex,kaastra2018_spex} format with \texttt{trafo} (v. 1.03). We used \texttt{trafo} to prepare a combined spectral file, in which each RGS1 and RGS2 first--order spectrum allocates a separate \texttt{SPEX} sector and region.
Such a combined spectral file was prepared for the analysis because it provides two clear advantages, as compared to analyzing the co--added data.
First, in the co-added spectra, observation--specific wavelength grid shifts at the detector plane can induce bias near the pixels and columns with abnormal response; as the X--ray flux from the source varies with time, artificial spectral features resembling emission and absorption lines may emerge in the co-added spectrum as a result (for more detailed description on the formation of such co-addition artifacts, see \citealt{kaastra11}). 
Second, the combined spectral file enables fitting the time--varying spectral components
independently for each observation, while at the same time the static components, 
such as absorption, can be fitted as a common value between the spectra.

In addition, we generated co--added RGS1 and RGS2 first order spectra (by combining the reduced
spectra with the SAS tool \texttt{rgscombine}), and a background--subtracted fluxed spectrum, 
in which all the RGS1 and RGS2 first-- and second--order data were combined to provide 
 the maximum signal--to--noise ratio (S/N) for the spectra (i.e., SAS \texttt{rgsfluxer} task generated spectra combined with the \texttt{SPEX} \texttt{rgsfluxcombine} tool). These two co--added spectra were generated for 
visual inspection only, not for analysis purposes, 
because of the co-addition artifacts that these spectra are likely to contain.

\subsection{FUV data}\label{fuv_red}

Ton~S~180 has been observed in the FUV with both the Far-Ultraviolet Spectroscopic Explorer (FUSE) and the
Space Telescope Imaging Spectrograph (STIS) on the Hubble Space Telescope (HST).
The individual STIS G140M exposures were reduced using the latest 
STIS pipeline software \texttt{CALSTIS} v. 3.4.2.
The flux--calibrated and individually--reduced exposures were co--added using inverse--variance weighting 
to improve the S/N. The final co--added spectrum covers the $1195-1300$~\AA\ range 
with a medium resolution of $\approx15000$ and with a S/N of $15-40$ per raw-pixel 
($\approx20$ per pixel around 1270 \AA, i.e., near the predicted wavelength of \ion{H}{I}~Ly$\alpha$ at $z_\mathrm{OVI}$).

The FUSE analysis was conducted using observation-level spectral files (OID's P1010502000 and D0280101000, taken 1999-12-12 and 2004-07-13, respectively), available at MAST\footnote{https://archive.stsci.edu/missions-and-data/fuse}. These data files contain fully calibrated, co--added spectra (i.e., combining the sub--exposures) for each FUSE segment and channel, processed with the final version (v. 3.4) of the CalFUSE data reduction pipeline \citep[for more details, see][]{Dixon_2007}. While three FUSE channels cover the wavelength band of the examined \ion{O}{VI} absorption (i.e., Lif1A, Sic1A and Sic2B), at this band, the Lif1A channel provides an about two times larger effective area than the two Sic detectors combined. Due to the low S/N ratio in the data of the Sic spectra, and the calibration issues that can follow from that, we consider only the Lif1A data in this work. In the analysis presented in Sect.~\ref{fuv_an}, we use co-added Lif1A spectra, where the two observations have been combined using inverse--variance weighting. This spectrum has a S/N of 5 per pixel in the band of interest.


\section{FUV analysis near z=0.04560}
\label{fuv_an}

As the basis for the search of X--ray absorption lines we begin by re--analyzing the FUV data for the previously detected absorption features near $z=0.0456$ \citep[e.g.,][]{danforth06,tilton12}. This reference redshift corresponds to that of 
the galaxy with the smallest impact parameter along the sightline near the absorber (as discussed in more detail in Sect.~\ref{disc}).


We analyzed the \ion{H}{I} Ly$\alpha$ lines in the STIS data and the \ion{H}{I} Ly$\beta$, \ion{C}{III} $977$, and \ion{O}{VI} $\lambda\lambda$ $1031.9$ lines in the FUSE data. The weaker \ion{O}{VI} $\lambda\lambda$ $1037.6$ line was not measured owing to the low S/N of the data. For each line, we normalized the data using low--order polynomial fits to surrounding line--free regions. We fit the lines with Voigt profiles convolved with the appropriate line-spread functions (the tabulated STIS LSF\footnote{http://www.stsci.edu/hst/stis/performance/spectral\_resolution} and an assumed Gaussian FUSE LSF with FWHM=$\SI{20}{km\per\s}$)
using \texttt{VPFIT}\footnote{https://www.ast.cam.ac.uk/~rfc/vpfit.html}. In each case, we used the minimum number of Voigt components required to produce a reduced  $\chi^2\sim1$. The resulting fits provide centroid wavelengths, line widths, and column densities from each transition.


\begin{table}[htb]
\caption{Measurements of FUV absorption lines near $z=0.0456$}
\begin{center}
\label{table:FUV}
\begin{tabular}{lcccc}
\hline
\hline
Line ID &  $z$ &  $\Delta v$  &  $b$ &  log$_{10} N$  \\
 &    & $(\mathrm{km\,s^{-1}})$ & $(\mathrm{km\,s^{-1}})$ & $N$ in $(\mathrm{cm^{-2}})$ \\
\hline

Ly$\alpha$ &  $0.04512$ & -138.2    &           $52.0\pm6.7$ & $13.46\pm0.04$   \\  
Ly$\beta$ &  $(0.04508)$ &                     & $16.9\pm14.8$ & $13.50\pm0.25$   \\
\hline
\ion{O}{VI} &  $0.04557$ & -8.5    &           $17.8\pm11.2$ & $13.46\pm0.16$ \\
\hline
Ly$\alpha$ &  $0.04567$ &   +18.9   &           $45.5\pm3.6$ & $13.70\pm0.02$   \\
Ly$\beta$ & $(0.04566)$ &                      & $27.6\pm10.1$ & $13.90\pm0.15$   \\
\ion{C}{III} & $(0.04565)$   &                 & $25.8\pm9.7$ & $13.08\pm0.12$   \\  
\hline
\ion{O}{VI} &  $0.04579$ & +54.8    &           $17.0\pm7.1$ & $13.68\pm0.10$ \\
Ly$\alpha$ & &                      &          $<88.5^*$ & $<12.87^{**}$ \\

\hline
\end{tabular}
\end{center}
\tablefoot{Absorbers near $z=0.04560$. The detected ions at each redshift are listed according
to descending detection significance. 
The horizontal lines group the absorption lines into different absorbers according to the measured wavelengths of the line features.
All the quoted redshifts have a statistical uncertainty of $\approx10^{-5}$. $\Delta v$ is the velocity offset with respect to $z=0.04560$. 
All Ly$\alpha$ measurements were made using STIS data, while the other lines were measured in the FUSE data.
\newline
$^*$ Predicted from the $b_\mathrm{OVI}$ measurement assuming pure thermal broadening.\newline
$^{**}$ $3\sigma$ upper limit when $b_\mathrm{HI}=\SI{88.5}{\km\per\s}$.}
\end{table}

\begin{figure}[ht]
\begin{center}
\includegraphics[width=9cm]{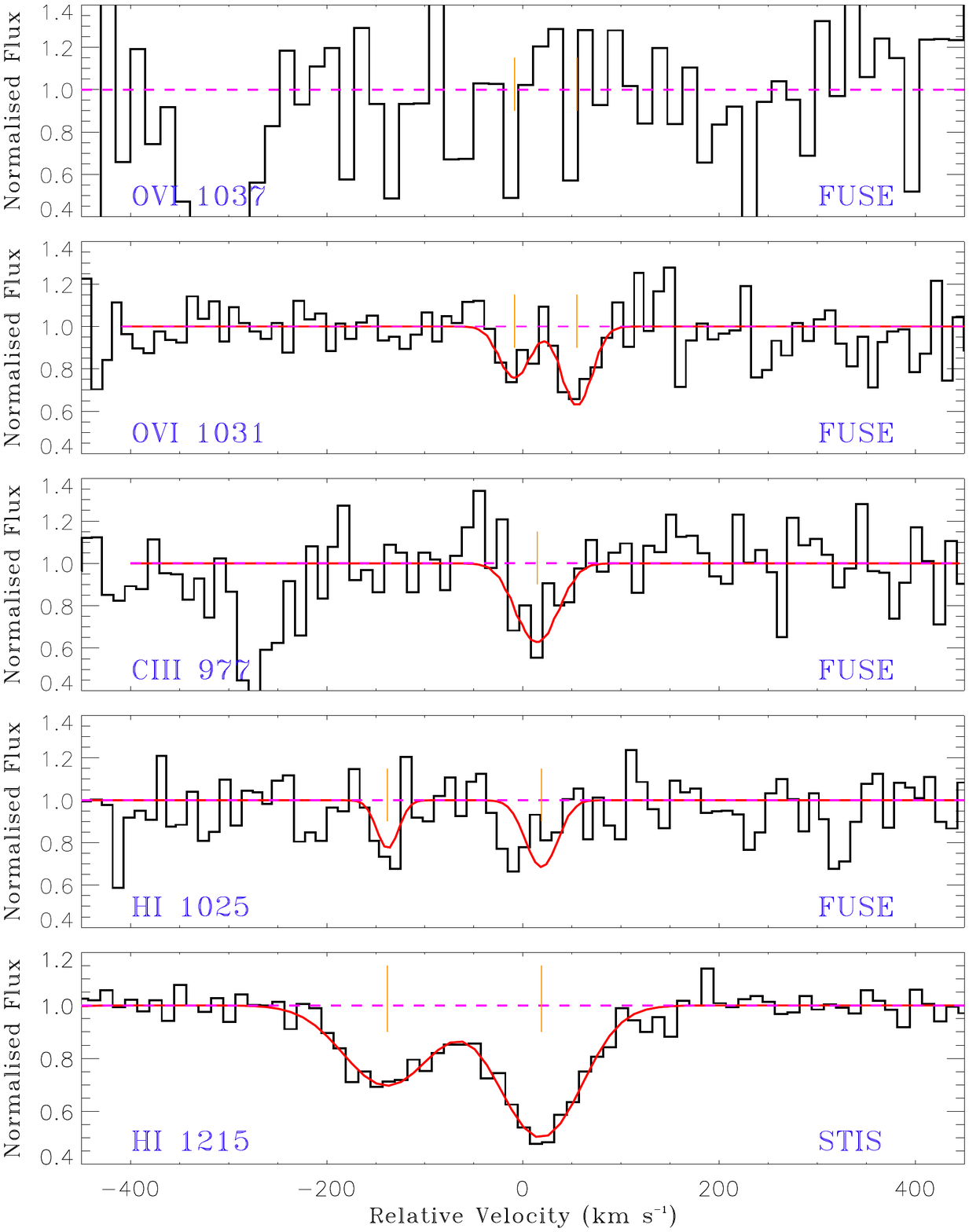}
\end{center}
\caption{FUSE and STIS spectral data of \ion{O}{VI}, \ion{C}{III}, \ion{H}{I} in velocity space relative to $z=0.04560$. The \ion{O}{VI} $\lambda$ $1037.6$ \AA\ line was not included in the fits due to the poor S/N at the relevant wavelengths (top panel). 
The unlabeled absorption feature blueward of the \ion{C}{III} lines is the Galactic \ion{Si}{II} $\lambda$ $1020$ \AA\ line. A binning factor of 3 has been used for FUSE data in this illustration.
}
\label{fig:FUV}
\end{figure}

The results of the best--fit models of the four absorbers closest to $z=0.04560$ are presented in
Table \ref{table:FUV}, and the \ion{O}{VI}, \ion{C}{III}, \ion{H}{I} best-fit models 
are shown in Fig.~\ref{fig:FUV}.
The $z=0.04579$ \ion{O}{VI} line yields a best-fit width, $b_\mathrm{OVI}=17.0\pm7.1$~$\SI{}{km\per s}$, which is roughly equal to the FUSE limit ($R\sim 15000$). 
A similar result is obtained for the weaker \ion{O}{VI} feature at $z=0.04557$, thus implying that both of these lines arise from warm/cool gas.

The best-fit centroid wavelength ($\lambda_{\rm OVI}\approx1079.18$~\AA) of the $z=0.04579$ \ion{O}{VI} line lies close to a Galactic
H$_2$ transition (L2R2 1079.2254 \AA). However, since the stronger $J=2$ rotational level transitions of H$_2$ 
(such as L3R2 1064.9948 \AA, L4R2 1051.4985 \AA, L5P2 1040.3672 \AA\ etc.) are not detected,
we conclude that the \ion{O}{vi} line is not
contaminated with Galactic H$_2$ absorption. 
There is no clear indication of \ion{H}{I} absorption associated with the $z=0.04579$ \ion{O}{vi} 
line; the velocity offset of the nearest \ion{H}{I}~Ly$\alpha$ line ($\Delta v\approx 35\rm~km~s^{-1}$) is larger than plausible intrumental uncertainties
\citep[as discussed e.g., in][]{danforth06}. The same is true for the weaker \ion{O}{VI} feature detected at $z=0.04557$ ($\Delta v\approx 27\rm~km~s^{-1}$).
We therefore assume that the \ion{O}{VI}--\ion{H}{I} misalignments are physical,
and that the Ly$\alpha$ lines associated with the detected \ion{O}{VI} absorbers
are not detected due to their low column densities (i.e., high gas metallicity), 
broad line profiles, fitting complications due to blending with the
$z=0.04567$ Ly$\alpha$ line, or some combination thereof. In contrast, the only detected \ion{C}{III} line is found to match the redshift of the $z=0.04567$ Ly$\alpha$ absorber. 
Our analysis of the FUV data is in general agreement with that of \cite{danforth06}, who had also found a statistically significant \ion{O}{VI} $\lambda$ 1031.9 \AA\ line at a similar redshift ($\Delta v\approx70$~km$\,\mathrm{s}^{-1}$ with respect to $z_\mathrm{HI}=0.04560$), with only marginal evidence of the associated 1037.6 \AA\ line.

\section{X-ray spectral modeling}\label{X-ray}

The X--ray spectral analysis of this work was conducted with \texttt{SPEX} vs. 3.03.00\footnote{http://www.sron.nl/SPEX} using \cite{lodders09} proto--Solar abundances and \cite{bryans09} collisional ionization equilibrium (CIE) ion fractions (i.e., the \texttt{SPEX} default tables). We used \texttt{SPEX} to rebin the data to 20 m\AA\, wavelength bins, which oversamples the RGS energy resolution by a factor of $\approx3$.
Due to the low number of spectral counts per channel in some of the observations, the spectral models were minimized using 
Cash statistics \citep{cash78}. 

We model the X--ray spectra using an absorbed power--law model 
and fit the data with a (partial) joint fit method. In these fits, the absorption component fit parameters are coupled across spectral observations, while the power--law continua, 
being highly time variable for Ton~S~180, are modeled independently for each
observation. The absorption model includes components for (1) 
the hot Galactic halo; (2) the neutral Galactic halo \footnote{http://www.swift.ac.uk/analysis/nhtot/index.php},
 whose $N_\mathrm{H}$ is free to vary between $1.3-1.7\times10^{20}$~cm$^{-2}$;
(3) the hot halo of the source galaxy; (4) 
a redshifted intergalactic absorber, which is the main component of interest
for this paper. 
Absorption components 1--3 are modeled using the \texttt{SPEX} `hot' components, that is, absorption through an optically thin layer of gas in collisional ionization equilibrium. In case of the neutral halo modeling, we fix the gas temperature to $kT=5\times10^{-4}$~keV, which is a parameter choice for the \texttt{SPEX} `neutral plasma' modeling. Other than that, we let the defining parameters of the continuum shapes (photon index and normalization parameter) and absorption components ($N_\mathrm{H}$ and $kT$) vary throughout the analysis. This ensures that the uncertainties of these model parameters are propagated correctly to the uncertainties of the model parameters of interest.

For the redshifted absorption component (4), which models the intervening WHIM absorption, two different model components were used. We begin the analysis by adopting a single--ion absorption model (\texttt{SPEX} `slab'), so as to examine oxygen absorption through the absorber. Then, we use the \texttt{SPEX} `hot' CIE absorption component in order to investigate the ionization conditions within the absorber.

\section{Results of the X--ray spectral analysis}

\subsection{Oxygen absorption at z=0.04579}\label{slab}

\ion{O}{VII} and \ion{O}{VIII} lines at $z_\mathrm{OVI}=0.04579$ were first examined individually with the redshifted `slab' model. We limit the fitting band to $16-24$~\AA\, which contains all the strongest
oxygen lines for the redshifted and the non-redshifted (Galactic) absorber components. The results of the `slab' analysis are presented in the second column of Table~\ref{table:main}. 

The spectral modeling of \ion{O}{VII} absorption yields log$\,N_\mathrm{OVII}(\mathrm{cm^{-2}})=16.52_{-0.28}^{+0.25}$, corresponding to a 
$5.0\sigma$ statistical significance for the detection of 
redshifted oxygen. 
The \ion{O}{VIII} signal is not significantly detected in the data, although the fit yields the best-fit value log$\,N_\mathrm{OVIII}(\mathrm{cm^{-2}})\approx15.7$. Whereas the best-fit predicted \ion{O}{VIII}~Ly$\alpha$ signal is not strong enough for visual identification at the S/N level of the individual spectra, we note that this signal does appear more prominent in the fluxed spectrum, where all the data of RGS1, RGS2 first and second spectral orders are combined to attain the maximal S/N spectrum for visual inspection (as is shown later in this paper).

\begin{figure}[ht]
\begin{center}
\includegraphics[width=3.9in]{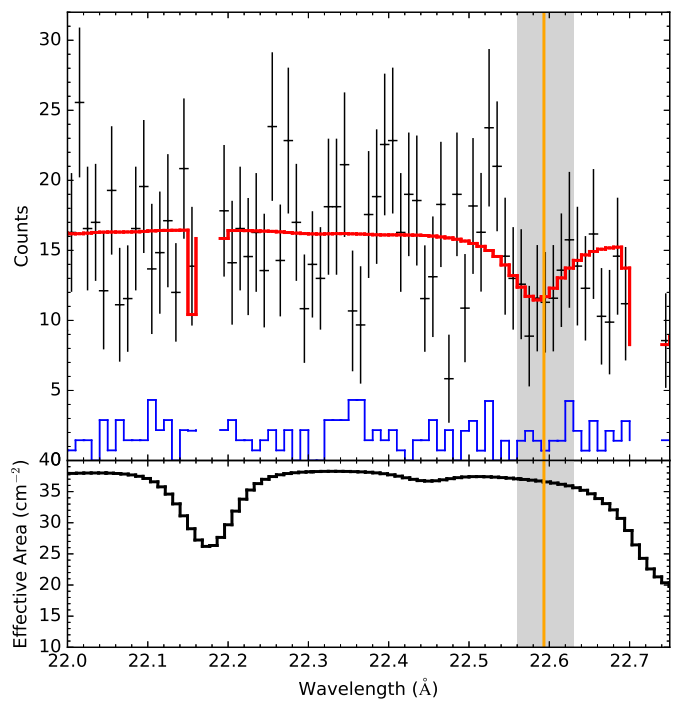}
\caption{Top panel: Zoom in on the non-binned count spectrum of the longest, $124$~ks RGS1 first--order exposure near the detected \ion{O}{VII}~He$\alpha$ line. The red curve is the best-fit of the redshifted SPEX `slab' model minimized to the data shown, yielding ion column log$N_{\mathrm{OVII}}$~(cm$^{-2}$)=$16.5$, i.e., a value consistent with the main analysis. The orange line marks the predicted \ion{O}{VII} line centroid and the blue curve the subtracted background in each channel. The missing data bins were removed by the reduction software due to the bad response (see Sect.~\ref{x-rayre} for details).
Below: The effective area of the RGS1 instrument in the corresponding wavelength range, showing two instrumental features.
}
\label{fig:area}
\end{center}
\end{figure}

We acknowledge that the RGS instruments contain a large number of instrumental features which can cause measurement bias if coinciding with the wavelength bins of a modeled spectral feature. To test the reliability of the redshifted \ion{O}{VII} measurements,
 we investigate the instrument effective area at the spectral location of the detected \ion{O}{VII}~He$\alpha$ line (Fig.~\ref{fig:area}). We find that the closest instrumental feature is located far enough from the modeled line profile not to affect the measurement, thus indicating that this absorption feature has an astrophysical origin.

\begin{table*}[ht]
\caption{X-ray and FUV measurements of the Ton~S~180 oxygen absorber}
\begin{center}
\label{table:main}
\small
\begin{tabular*}{0.75\linewidth}{l|c|c|c|c|c}
\hline
\hline

\multicolumn{1}{c|}{} &  \multicolumn{4}{c|}{X-RAY} &  \multicolumn{1}{c}{FUV}  \\

\multicolumn{1}{c|}{} &  \multicolumn{2}{c|}{$z=0.04579$} & \multicolumn{2}{c|}{$z=0.0455$} & \multicolumn{1}{c}{$z=0.04579$}  \\

\multicolumn{1}{l|}{parameter} & \multicolumn{1}{c|}{SLAB}  & \multicolumn{1}{c|}{CIE} & \multicolumn{1}{c|}{SLAB}  & \multicolumn{1}{c|}{CIE} & \multicolumn{1}{c}{}  \\

\hline

log $N_{\mathrm{OVI}}$ (cm$^{-2}$) & - & 13.8$\pm0.2$  &	- & 13.9$\pm0.2$  & 13.68$\pm0.10$ \\

log $N_{\mathrm{OVII}}$ (cm$^{-2}$) & $16.52_{-0.28}^{+0.25}$ & 16.4$\pm0.2$ &  	$16.69_{-0.39}^{+0.37}$ & 16.4$\pm0.2$ & - \\ 

log $N_{\mathrm{OVIII}}$ (cm$^{-2}$) & $15.7_{-1.1}^{+0.4}$ &  15.9$\pm0.2$ &  	$15.7_{-1.5}^{+0.4}$ &  16.0$\pm0.2$ &- \\ 

$\sigma_\mathrm{OVI}$  & - & - & 							- & - & 5.4 \\

$\sigma_\mathrm{OVII}$  & 5.0 & - & 							 5.1 & - &- \\

$\sigma_\mathrm{OVIII}$  & (1.1) & -  &							(1.0)   & -  \\

$b_\mathrm{tot,oxygen}$~(km$\,$s$^{-1}$)  & 141$\,^*$ & -  &	$106_{-42}^{+73}$  & - & $<38.1$  \\
$b_\mathrm{nt}$~(km$\,$s$^{-1}$)  & - & 141$\,^*$  &	-  & 97$\,^*$ & -  \\

$kT$ (kev) & - &  $0.14\pm 0.02$ & 						- &  $0.14\pm 0.02$ & $<0.12$  \\
$T$ (K) & - &  $1.7\pm0.2\times10^6$ & 							- &  $1.7\pm0.2\times10^6$ &$<1.4\times10^6$  \\

$N_\mathrm{H}$ ($Z_\sun/Z\times 10^{19}$ cm$^{-2}$)  & - & $5.3_{-2.1}^{+2.3}$ & 	- & $5.8_{-2.2}^{+3.0}$ &- \\ 
log $N_{\mathrm{HI}}^\mathrm{hot}$ (cm$^{-2}$)  & - & $12.9\pm0.2\,^{**}$ & 				- & $12.9\pm0.2\,^{**}$ &$<13.01\,^{***}$ \\ 

\hline
$\Delta$C & 12 & 15 & 14 & 15 & - \\
\hline
\end{tabular*}
\end{center}
\tablefoot{Results of the X--ray analysis with `slab' (for the
individual oxygen ions) and CIE absorption models, conducted on the fixed FUV redshift $z_\mathrm{OVI}=0.04579$
absorber (columns 2-3) and at the X--ray fitted hot absorber redshift $z_\mathrm{OVII}=0.0455$ 
(cols. 4-5), and the FUV analysis at the redshift of the stronger \ion{O}{VI} absorber (col. 6). 
The X--ray results on \ion{O}{VI} and \ion{H}{I} are model predictions. The quoted temperatures are the CIE ionization temperature (X--ray) and the upper limit on the \ion{O}{VI} ion temperature (FUV). The parameter values are quoted with their respective $1\sigma$ uncertainties, whereas the upper limits are given at their $3\sigma$ limit. $\Delta$C is the improvement of the fit statistics due to the addition of the redshifted absorption component (in SLAB columns, the \ion{O}{VII} `slab'). 
All the X--ray models yield C-stat/d.o.f.$\,\approx1.04-1.06$. 
\newline
\small{$\,^*$ Parameter was fixed during the minimization.\newline
$^{**}$ Calculated for $Z=Z_\sun$\newline
$^{***}$ Formal $3\sigma$ upper limit obtained from STIS assuming $b_\mathrm{HI}=\SI{200}{\km\per\s}$. See Sect.~\ref{interpretation} for details.}}
\end{table*}

\subsection{CIE absorption at z=0.04579}\label{cie}

The high \ion{O}{VII} column density measured with the `slab' model is a possible indicator of CIE conditions for the absorbing gas 
(see e.g., Fig.~11 in \citealt{wijers19}).
We examine this scenario by 
replacing the redshifted `slab' component with a 
redshifted CIE absorption component with proto--Solar abundances, 
and we fit the model parameters $N_\mathrm{H}(Z)$ and $kT_\mathrm{CIE}$. 
For this fit, we expand the fitting band to cover the wavelengths between $13-30$~\AA, 
so as to include additional model--constraining lines in the analysis,
including those of \ion{Ne}{ix}, \ion{N}{VII}, \ion{C}{VI}, and others. 
The results of the CIE modeling are presented in the third column of Table~\ref{table:main}.

The CIE modeling yields tight constraints on the gas temperature, $kT_\mathrm{CIE}=0.14\pm0.02$~keV ($T_\mathrm{CIE}=1.7\pm0.2\times10^6$~K) with the model scaling parameter $N_\mathrm{H}=5.3_{-2.1}^{+2.3}\times Z_\sun/Z_\mathrm{abs} \times10^{19} $~cm$^{-2}$.
The best--fit CIE model also indicates that:

(1) the CIE \ion{O}{VII} and \ion{O}{VIII} ion column densities agree with the `slab' results;  
(2) the CIE model predicts no other detectable lines in the RGS energy band for the available S/N level, and 
(3) the CIE model predicts $N_\mathrm{OVI}$ that matches the FUV measured \ion{O}{VI} column associated with the $z=0.04579$ \ion{O}{VI} line. 

The \ion{O}{VI} column-density predictions of the CIE model seemingly imply
that the FUV detected \ion{O}{VI} signal at $z_{\rm{OVI}}=0.04579$ is the counterpart of
the hot CIE phase, regardless of what other phases might be present in the absorbing medium.
Considering additional information provided by the FUV and X--ray measurements, we examine whether this interpretation is plausible. Namely, if all 
line detections originate from a common CIE phase, the \ion{O}{VI} line profiles should have the following broadening:
\begin{equation}\label{eq:b}
b_\mathrm{OVI}^\mathrm{CIE}= \sqrt{\frac{2kT_\mathrm{CIE}}{m}+b_\mathrm{nt}^2}\gtrsim42~\mathrm{km}\,\mathrm{s}^{-1},
\end{equation}
where $b_\mathrm{OVI}^\mathrm{CIE}$ is the predicted Doppler spread of \ion{O}{VI} lines, $m$ the atomic mass of oxygen and $b_\mathrm{nt}$ the nonthermal velocity dispersion ($b_\mathrm{nt}\geq0$).

In contrast, the FUV analysis yields $b_\mathrm{OVI}=17.0\pm7.1$~km$\,$s$^{-1}$ (Table \ref{table:FUV}), or a $3\sigma$ upper limit $b_\mathrm{OVI}<38.3$~km$\,$s$^{-1}$. Assuming the detected \ion{O}{VI} line is only thermally broadened, the $3\sigma$ upper limit on the \ion{O}{VI} ion temperature would correspond to $T_\mathrm{OVI}<1.4\times10^6$~K. 
The information on the line--widths therefore indicates that the FUV and X--ray measurements are unlikely to trace oxygen belonging in the same gas phase.

\section{Properties of
the FUV/X--ray absorbing medium towards Ton~S~180}
\label{interpretation}

In Sect.~\ref{cie} we showed that while the X--ray CIE absorption modeling yielded \ion{O}{VI-VIII}
column density constraints matching their freely fitted
ion column densities, the FUV measurement on the 
\ion{O}{vi} line width excludes the possibility of 
a single--temperature CIE plasma. We note that the disagreement
between the obtained $T_\mathrm{CIE}$ and $T_\mathrm{OVI}$ cannot be due to transient differences between the electron 
and ion temperatures in collisionally ionized heating/cooling gas because in low-density ($n_\mathrm{H}\lesssim10^{-4}$~cm$^{-3}$, $Z<Z_\sun$), heating gas, $\sim10^6$~K equipartition is reached faster than the ionization equilibrium due to electron collisions.
Similarly, recombination--lag--driven over-ionization 
of cooling gas \citep[e.g.,][and references therein]{yoshikawa06} seems implausible, 
given that the oxygen ion cooling times exceed the Hubble time 
 for typical 
WHIM metal number densities \citep[as is shown, e.g., in][]{bykov08}.

Alternatively, the absorber could be predominantly in photo--ionization equilibrium (PIE), at a
gas temperature within the measured limits of the \ion{O}{VI} ion temperature (for which the best--fit value of $b$ predicts $T_\mathrm{OVI}\approx3\times10^5$~K if the line is only thermally broadened). 
We therefore investigate the WHIM ion mass distributions in  \cite{khabibullin19}, which is based on the \texttt{magneticum}\footnote{www.magneticum.org} cosmological, hydrodynamical simulation. 
Their results show that, in principle 
$\sim10^6$~K CIE--like oxygen ionization conditions can occur in $T\sim10^5$~K WHIM in PIE.
Nevertheless, we find that in the case of the examined absorber, the measurement limits on $N_\mathrm{OVI}$, $N_\mathrm{OVII}$, $N_\mathrm{OVIII}$  
are inconsistent with the relative abundances of \ion{O}{VI}, \ion{O}{VII} and \ion{O}{VIII} in their simulations, 
as significant discrepancies remain even at the most closely matching parameter region ($T\approx4.5\times10^5$~K, $n_\mathrm{H}\sim10^{-5}$~cm$^{-3}$).
Considering our measurement constraints on \ion{O}{VI-VIII}, we find that the best match with the \cite{khabibullin19} 
results occurs at $T\approx1-2\times10^6$~K and $n_\mathrm{H}\approx2\times10^{-5}$~cm$^{-3}$, which is a solution located in the CIE dominated branch of their WHIM distribution.

\begin{figure}[ht]
\begin{center}
\includegraphics[width=3.8in]{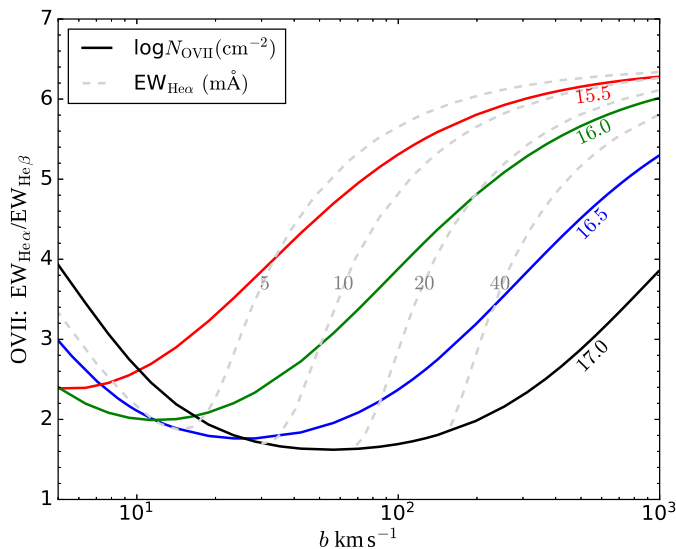}
\end{center}
\caption{
Theoretical \ion{O}{VII} He$\alpha$ and $\beta$ line equivalent width ratio as a function of line width $b$. The solid curves show the EW$_{\mathrm{He}\alpha}/$EW$_{\mathrm{He}\beta}$ -ratio for various \ion{O}{VII} ion column densities, as labeled. The dashed lines show a selection of constant EW$_{\mathrm{He}\alpha}$ curves in this plane, excluding the parameter regions where the associated ion column densities exceed $N_\mathrm{OVII}=10^{17}$~cm$^{-2}$. When moving, at fixed column density (i.e., along a solid curve), from high to low $b$--values, the equivalent width ratio decreases gradually from its optically thin value, due to the stronger saturation of the He$\alpha$ line relative to that of the weaker $\beta$ line. When He$\alpha$ line saturation reaches the damped regime, the opposite behaviour takes place.
These properties suggest that the EW$_{\mathrm{He}\alpha}/$EW$_{\mathrm{He}\beta}$ -ratio can be utilized as a sensitive diagnostic tool for $b$ in the range of X--ray detectable \ion{O}{VII} ion column densities, especially for low line-of-sight gas velocity dispersions. 
}
\label{fig:oviiSaturation}
\end{figure}

\subsection{Two--phase model}\label{two-phase}

Because the one--phase hypothesis seems like an improbable interpretation for the \ion{O}{VII}, \ion{O}{VI} measurements, we instead consider more complex thermal structures by investigating the correlation between the spatial
occurrence of warm and hot WHIM absorbers 
in the \texttt{EAGLE} simulation \citep[][]{wijers19}. 
We focus on a two--temperature scenario in which the X--ray signal arises from a hot CIE phase, whereas the FUV absorber traces warm gas,
which may be predominantly photo-- or collisionally--ionized, or in a mixed state. 

Since the FUV and X--ray detected gas phases are unlikely to be strictly co--spatial, 
we make use of the well--sampled \ion{O}{VII} signal and fit $z_\mathrm{OVII}$ with the \texttt{SPEX} `slab' model. With this fit we obtain $z_\mathrm{OVII}=0.0455\pm0.0005$, corresponding to $\Delta v=-16\pm140\,\SI{}{\km\per\s}$ with respect to $z=0.0456$. The best-fit redshift of the \ion{O}{VII} absorption is therefore consistent with both of the detected \ion{O}{VI} absorbers
 (i.e., $z_\mathrm{OVI}=0.04579$, $\Delta v=61.8\,\SI{}{\km\per\s}$ and $z_\mathrm{OVI,weak}=0.04557$, $\Delta v=-8.5\,\SI{}{\km\per\s}$) within the measurement uncertainty.

In addition, we note that the strong \ion{O}{VII}~He$\alpha$ signal can be used to constrain $b_\mathrm{OVII}$, even if the 
RGS resolving power does not allow direct measurements of the \ion{O}{VII} line widths. This is because at the X--ray--detectable ion column densities ($N_\mathrm{OVII}\gtrsim 10^{15}$~cm${^{-2}}$), the strongest \ion{O}{VII} lines will be saturated at the line center to a degree that depends on the \ion{O}{VII} ion column density and the line--of--sight velocity dispersion of the absorbing ions.
In Fig.~\ref{fig:oviiSaturation} we demonstrate the effects of line saturation on observations of X--ray detectable \ion{O}{VII} column densities in the WHIM regime.
The constant EW$_\mathrm{He\alpha}$ (dashed) curves show that the ratio between the lower limit on EW$_\mathrm{He\alpha}$ and the upper limit on EW$_\mathrm{He\beta}$, constrain the lower limit on $b_\mathrm{OVII}$, and vice versa.
Indeed, the lower limit on $b_\mathrm{OVII}$, being dependent on the lower limit of the stronger and 
the upper limit of the weaker line, is always obtainable when the \ion{O}{VII}~He$\alpha$ line is significantly detected. In contrast, to place an upper limit on $b_\mathrm{OVII}$, a detection of the He$\beta$ line is also required.
Furthermore, the figure illustrates that the narrow line widths (e.g., $b_\mathrm{OVII}<\SI{100}{km\per\s}$), which are most common in \ion{O}{VI} absorbers, require higher column densities of \ion{O}{VII} for X--ray detection than the broader lines require (as indicated by the crossings of the EW$_\mathrm{He\alpha}$ curves over the $N_{\rm OVII}$ curves). It is noteworthy that this scenario contrasts with the limitations of FUV \ion{O}{VI} searches, which are most sensitive to narrow lines \citep[e.g.,][]{richter06b}.

Therefore, we free the \texttt{SPEX} `slab' velocity dispersion parameter in the modeling, so that the free parameters of the redshifted absorption component include $N_\mathrm{OVII}$, $b_\mathrm{OVII}$ and $z_\mathrm{OVII}$. This fit yields $b_\mathrm{OVII}=106_\mathrm{-42}^{+73}$~km$\,$s$^{-1}$, which exceeds the thermal broadening for oxygen lines calculated with Eq.~\ref{eq:b} ($b_\mathrm{th}\approx\SI{42}{\km\per\s}$), and indicates a non--thermal broadening of $b_\mathrm{nt}=97_{-49}^{+77}$~km$\,$s$^{-1}$. 
We find that adopting this smaller $b_\mathrm{OVII}$--value would
increase the statistical significance of the \ion{O}{VII} detection from $5.0\sigma$ to $6.7\sigma$, thus implying that the \ion{O}{VII} line ratio is affected by stronger saturation than what is predicted if one adopts the \texttt{SPEX} default for the line-of-sight velocity dispersion, which corresponds to $b=\SI{141}{\km\per\s}$. 

\begin{figure}[ht!]
\begin{center}
\includegraphics[width=0.86\columnwidth]{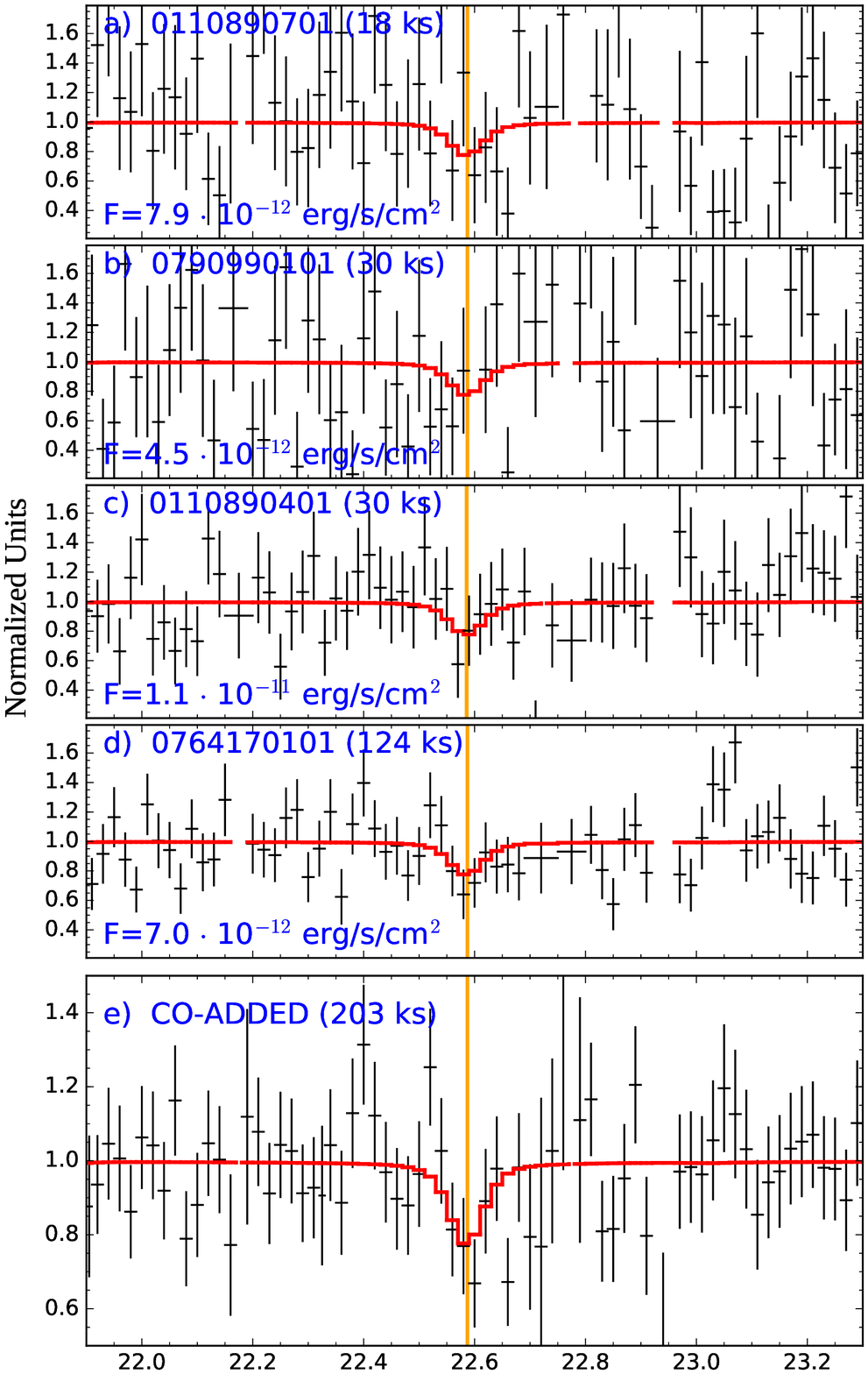}
\includegraphics[width=0.86\columnwidth]{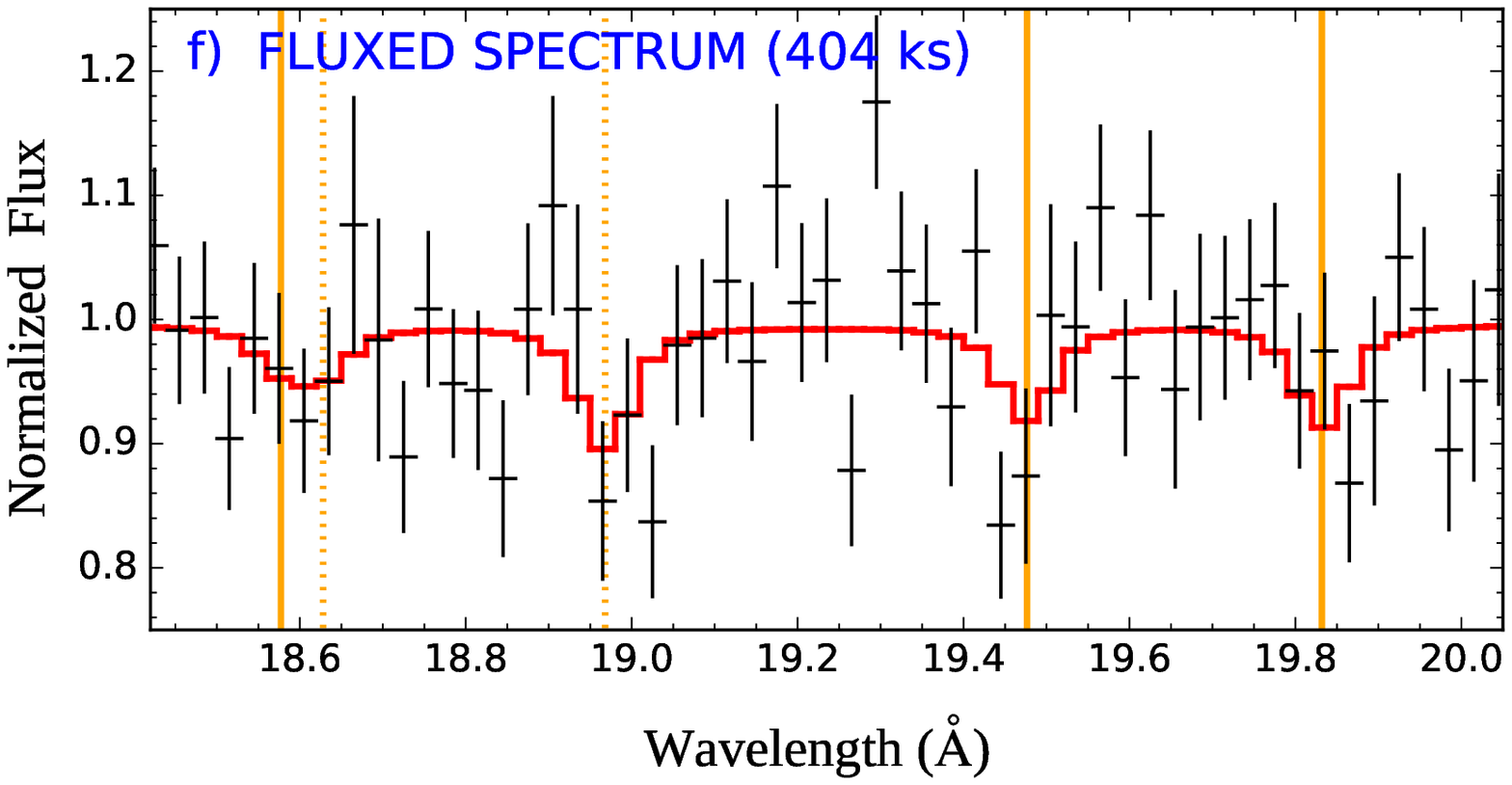}
\end{center}
\caption{Panels a -- d: Continuum normalized RGS1 1st order spectra of the reduced observations (binning factor 2), zoomed in around the $z=0.0455$ OVII He$\alpha$ wavelength, indicated by the orange line. The red curve is the best-fit model including the Galactic and redshifted \texttt{SPEX} `hot' absorption components (Sects. \ref{X-ray}, \ref{two-phase}). These components were jointly--fitted to the individual spectra (i.e., RGS1--RGS2 1st), whereas the emission continua were fitted independently for the data of each observation. The clean times and the source fluxes (0.5--2 keV) are quoted. Due to the failed RGS2 instrument CCD array \#4, and zero effective area of higher dispersion orders above $\lambda\gtrsim20$~\AA, only RGS1 1st order data covers this spectral band. Panel e: Co-added spectrum of the RGS1 1st data at the same band. The absorption model is the same as in panels a--d, but the continuum was fitted to the co-added data. Panel f: A fluxed spectrum (binning factor 3) combining all the reduced RGS data (i.e., RGS1--2 1st and 2nd dispersion orders) at the band containing a set of weaker spectral lines produced by the same best-fit absorption model as above. The centroid wavelengths of the $z=0.0455$ lines are marked with solid, and the Galactic lines with dotted, lines. From left to right they are: redshifted \ion{O}{VII} He$\gamma$, Galactic \ion{O}{VII} He$\beta$, 
Galactic \ion{O}{VIII} Ly$\alpha$, redshifted \ion{O}{VII} He$\beta$, redshifted \ion{O}{VIII} Ly$\alpha$. The continuum was fitted to the fluxed spectrum.}
\label{fig:oviib}
\end{figure}

To complete the X--ray analysis, we utilize the information from the \ion{O}{VII} `slab' modeling and re-fit the redshifted CIE model and \ion{O}{VIII} `slab' while fixing the redshift and non-thermal velocity dispersion to the obtained best-fit values, since one expects these parameters to be shared with all the ionic lines originating from the same gas phase.
The results of these fits are presented in columns 4 and 5 of Table~\ref{table:main}. The X--ray data and the best-fit CIE absorption model is presented in Fig.~\ref{fig:oviib}.

\subsection{Detectability of FUV lines associated with the hot phase}\label{non-detection}

\begin{table*}[ht]
\caption{Posterior probabilities of the Bayesian analysis at the redshift of the oxygen absorber.}
\begin{center}
\label{table:bayes}
\small
\begin{tabular*}{\linewidth}{l|c|c|c|c|c}
\hline
\hline

\multicolumn{1}{c|}{} &  \multicolumn{2}{c|}{SINGLE PHASE} & \multicolumn{2}{c|}{TWO PHASE} &  \multicolumn{1}{c}{X-RAY}\\

\multicolumn{1}{l|}{parameter} & \multicolumn{1}{c|}{MODEL A}  & \multicolumn{1}{c|}{MODEL B} & \multicolumn{1}{c|}{MODEL C}  & \multicolumn{1}{c|}{MODEL D} & \multicolumn{1}{c}{PREDICTION} \\

\hline

$b_\mathrm{OVI}$~(km$\,$s$^{-1}$)  & $13.1_{-10.4}^{+18.1}$ & $14.9_{-8.1}^{+12.4}$ & $13.2_{-10.6}^{+19.3}$ & $14.6_{-11.7}^{+26.3}$ & - \\

log $N_{\mathrm{OVI}}$ (cm$^{-2}$) & $13.66_{-0.16}^{+0.58}$ & $13.59_{-0.15}^{+0.13}$ & $13.66_{-0.17}^{+0.59}$ & $13.64_{-0.24}^{+0.48}$ & - \\

$z$ & $0.04578\pm 2\times 10^{-5}$ & $0.04578\pm 2\times 10^{-5}$ & $0.04578\pm2\times 10^{-5}$ & $0.04578\pm 2\times 10^{-5}$ & - \\

$b_\mathrm{OVI}$~(km$\,$s$^{-1}$), broad comp.  & - & - & 106$\,^*$  & $105.0_{-51.2}^{+45.7}$ & $106_{-42}^{+73}$ \\

log $N_{\mathrm{OVI}}$ (cm$^{-2}$), broad comp. & - & -  & $<13.98\,^{**}$ & $<15.00\,^{**}$ & $13.9\pm 0.2$\\

$z$, broad comp.  & - & - & $0.0456_{-0.0005}^{+0.0007}$ &  $0.0457\pm 0.0006$ & $0.0455\pm 0.0005$ \\

\hline

\end{tabular*}
\end{center}
\tablefoot{Results of the Bayesian analysis for models A--D, as defined in Sect.~\ref{bayes_method} (first 4 columns). 
X--ray analysis predictions for the broad \ion{o}{vi} line (counterpart to \ion{o}{vii}) are listed in the last column. The best--fit values and their 68\%-confidence quantiles are quoted. 
\newline
\small{$\,^*$ Parameter was fixed during the analysis.\newline
$^{**}$ 99.73\%-confidence upper limit. }
}
\end{table*}

\subsubsection{\ion{O}{VI}}\label{bayes_method}

The CIE model described in Sect.~\ref{two-phase} predicts $b_\mathrm{OVI}=106_{-42}^{+73}$~km$\,$s$^{-1}$ and log~$N_\mathrm{OVI}$~(cm$^{-2})=13.9\pm0.2$ for the non-detected, hot phase \ion{O}{VI} lines. Such a shallow, broad line profile would be at the limits of detetectability in the FUSE data, and its detection is degenerate with the continuum model (Fig.~\ref{fig:FUV}). Additionally, the predicted absorption signal would overlap with the narrower feature from the cooler gas phase, so any model of or limits on the hot phase \ion{O}{VI} signal must be determined jointly with the properties of the narrow line and the continuum.

To better understand the range of plausible absorption parameters for both the narrow features and the possible, predicted broad feature, we sampled the posterior probability distribution of the absorption and continuum parameters with version 3.0.2 of \texttt{emcee} \citep{foreman13}, which implements the affine-invariant Markov chain Monte Carlo (MCMC) ensemble sampler from \citep{goodman10}. This Bayesian approach allowed us to consider several combinations of models and prior probability distributions, while avoiding assumptions of Gaussianity in the posterior distribution. In all cases, we optimized the likelihood function using 1000 walkers initialized randomly within the domain. We discarded a number of initial steps equal to twice the maximum autocorrelation time in a parameter as a burn-in period, and we thinned the resulting chains of samples by a factor of half an autocorrelation time to reduce autocorrelation within each chain of samples.

The analysis was conducted using the observation--specific spectral files separately.
Though we considered several \ion{O}{VI} absorption models with various assumed prior probabilities, we always performed a joint fit to the two available FUSE observations.
We assumed a local linear continuum model with a slope that is fixed across the observations, but with different normalizations for the two observations owing to the different flux levels of the background AGN on the two dates. In all subsequent discussion of the absorption parameters, we have marginalized over the posterior continuum parameter distributions.

\begin{figure}[ht!]
\begin{center}
\includegraphics[width=0.9\columnwidth]{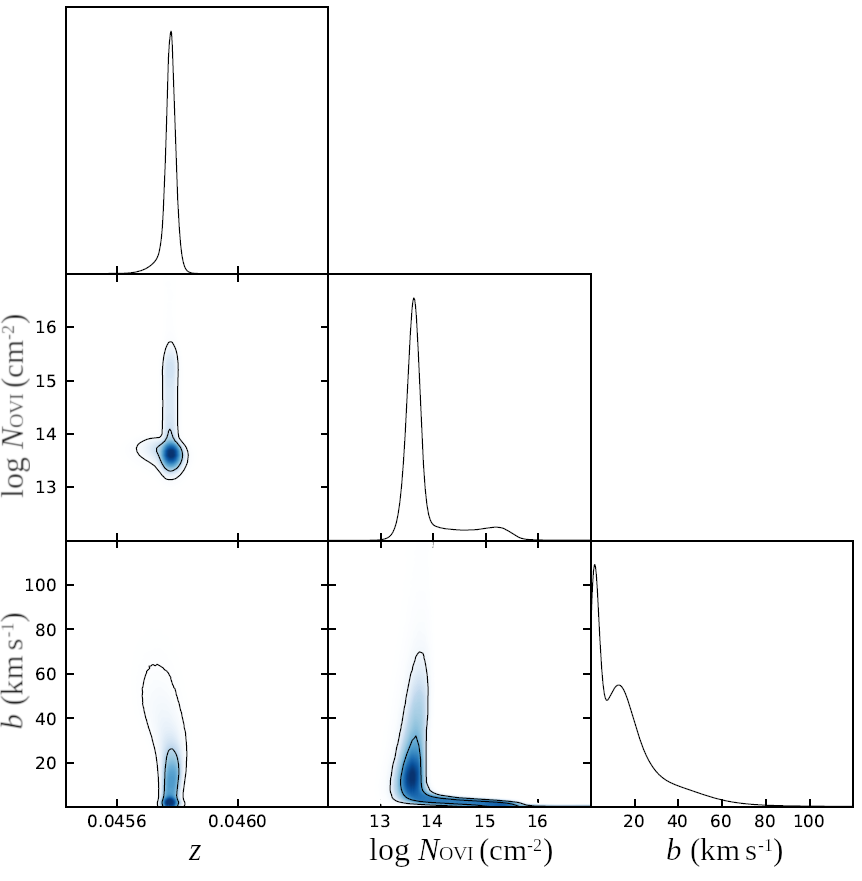}
\end{center}
\caption{Posterior probability distributions for the parameters of the narrow \ion{O}{VI} component obtained from the assumptions of MODEL A, marginalizing over the continuum parameters. We note that the allowed redshifts and line widths are incompatible with the expectations from the X--ray analysis.}
\label{fig:ovicorner}
\end{figure}

We first consider a simple single line model to investigate the posterior probability distributions of  $N_\mathrm{OVI}$, $z$ and $b$ over the FUV \ion{O}{VI} absorber. Adopting uniform priors over log~$N_\mathrm{OVI}$~(cm$^{-2})<17$, $z=0.0458\pm0.0005$ and $b=0-120~\mathrm{km\,s^{-1}}$, this approach yields the median parameter values and 68\%-confidence quantiles summarized as MODEL A in Table~\ref{table:bayes}. The posterior probability distributions for this model are shown in Fig.~\ref{fig:ovicorner}.

\begin{figure}[ht!]
\begin{center}
\includegraphics[width=0.9\columnwidth]{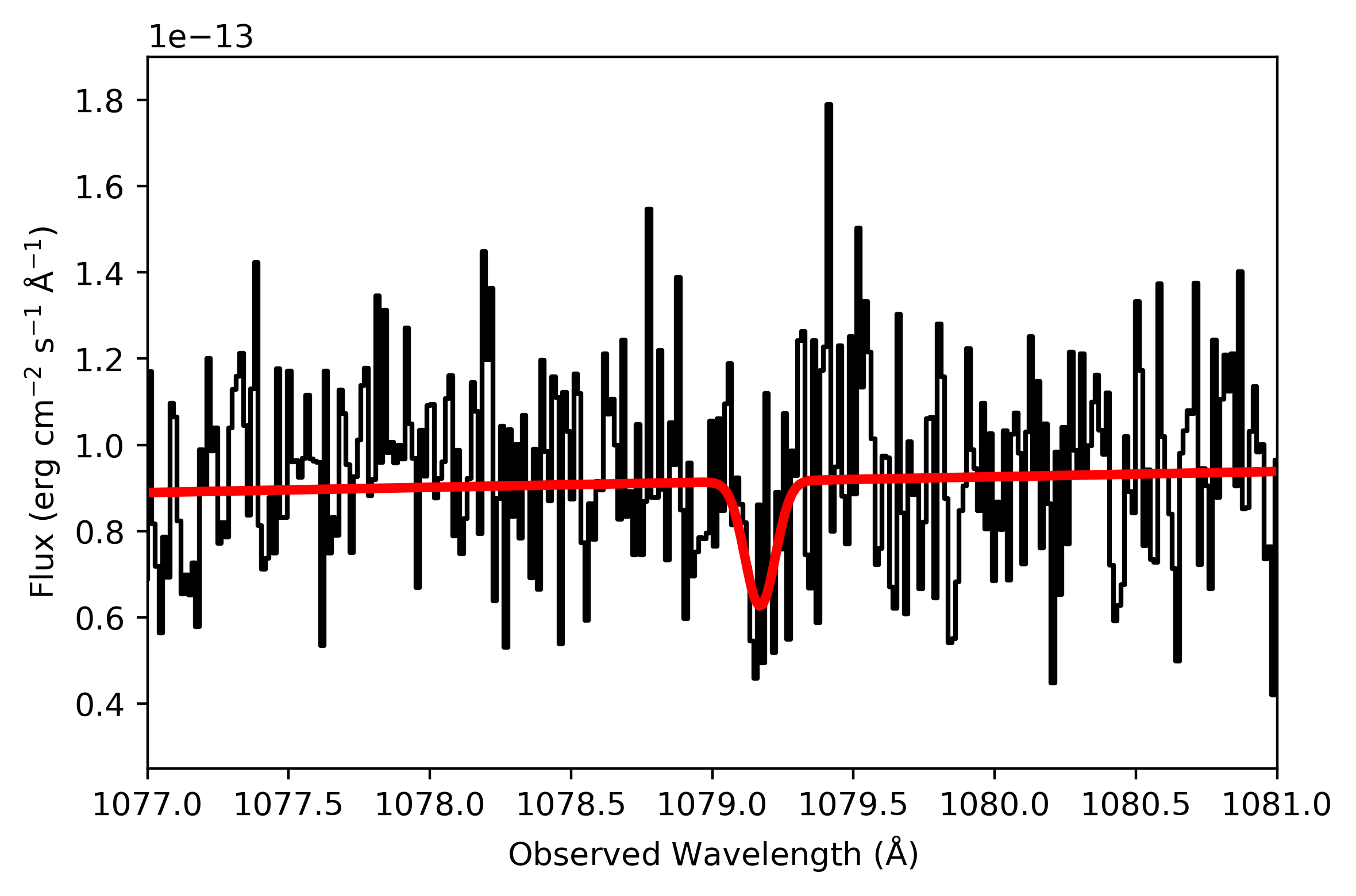}
\end{center}
\caption{Absorption model using the median parameter values from MODEL B, shown overplotted on the 1999 FUSE data, which is the observation with the highest signal-to-noise ratio. The median values from the other models are not shown, as they qualitatively appear almost identical to this one.}
\label{fig:ovifit}
\end{figure}

However, we know from surveys that IGM absorbers are not uniformly probable in strength or width, and it might be reasonable to adopt more stringent priors based upon these intergalactic \ion{O}{VI} absorber surveys. The MODEL B column of Table~\ref{table:bayes} summarizes the results of such a model, in which the prior on $z$ remains uniform, but the other line parameters use priors taken from IGM surveys. The prior line width distribution above $b=28~\mathrm{km\,s^{-1}}$ is described by Gaussian with $\mu=28$~km$\,$s$^{-1}$ and $\sigma=16\rm~km/s$, as in Fig.~7 of \cite{tilton12}. Below $b=28$~km$\,$s$^{-1}$, the probability is assumed to be uniform,  owing to the incompleteness in the IGM survey arising from line blending and spectral resolution. The log$\,N_{\mathrm{OVI}}$ prior probability distribution is constructed such that above log~$N_\mathrm{OVI}$~(cm$^{-2})=13$, the probability falls off according to a broken power-law distribution as described in Sect.~4.2.1 in \cite{danforth16}. Below log~$N_\mathrm{OVI}$~(cm$^{-2})=13$, the probability is assumed to be uniform, again owing to the incompleteness of the IGM survey. We note that these are rough approximations of our prior knowledge: they  neither account for covariances between the parameters, nor do they address potential incompleteness at the high-$b$ end of the distribution, which might arise, for example, from uncertain continuum models in the surveys. 
The model given by the median parameter values is plotted in Fig.~\ref{fig:ovifit}.

These two single-phase models yield similar results and illustrate that the qualitative results of our single-component fit are fairly robust to our choice of priors. While the most likely column densities are compatible with that predicted by our X--ray analysis, the posterior distributions of the redshift and $b$-parameter imply that this \ion{O}{VI} absorption component is unlikely to trace the same hot phase as the \ion{O}{VII} absorption. Specifically, we find that $90.9$~\% (MODEL A) and $96.2$~\% (MODEL B) of the posterior probability falls below $b=42$~km$\,$s$^{-1}$, which is smallest width expected of a line arising from a hot phase (Eq.~\ref{eq:b}). Obviously, considering the $b_{\rm{OVI}}$ constraints arising from the X--ray EW -measurements, a one--phase solution appears increasingly less likely.

Because a single \ion{O}{VI} absorption component yields results that are incompatible with the X--ray measurements, we also investigated a two-component model to determine if the FUSE data is compatible with an undetected, broader \ion{O}{VI} component that could arise from the same gas as the detected \ion{O}{VII} line. We considered a simplistic two-phase model with uniform priors on the parameters of a narrow component and the column density of a second component (as in MODEL A), but with the $b$-parameter of the broad component fixed at 106~km$\,$s$^{-1}$ (as in the best-fit X--ray model) and its redshift prior described by a Gaussian distribution, with mean at $z=0.0455$ (as in the best-fit X--ray model) and a variance that includes the X--ray measurement uncertainty, FUSE wavelength calibration uncertainty, and the X--ray wavelength uncertainty added in quadrature. The MODEL C column of Table~\ref{table:bayes} summarizes the results of this model.

If we further allow the line width of the broad component to vary by adopting a Gaussian prior on $b$ centered on the best-fit X--ray value a variance corresponding to the X--ray measurement's lower $1\sigma$ uncertainty, then we obtain the results given in the MODEL D column of Table~\ref{table:bayes}. As we might expect, this choice of priors weakens the upper limits that the model places on the column density, because broader lines allow higher column densities to be hidden by the noisy continuum. We note that we did not attempt to reproduce the asymmetric uncertainty distribution obtained in the X--ray measurement (due to the measurement technique's higher sensitivity towards low $b$--values, see Sect.~\ref{two-phase}), because at extremely high $b$-values (e.g., $b\gg150$~km$\,$s$^{-1}$), the line profile both reaches unphysically large values and is highly degenerate with the continuum parameters.

Regardless of our choice of priors, the width and redshift of the broader component in these two--phase fits are largely unconstrained by the data, and instead are simply tracing our choice of priors, as we would expect in the case of a non-detection. However, the analysis does provide useful upper limits on the column density of gas. These limits indicate that a broad \ion{O}{VI} line would be undetectable over a major fraction of the X--ray constrained parameter space, given the photon--statistics of the available FUV data. 
In particular, we find the 
data is consistent with a broad line profile centered at $z\approx0.0455-0.0458$ and characterized by $b\sim100$~km$\,$s$^{-1}$ and log~$N_\mathrm{OVI}$~(cm$^{-2})\sim13 - 14$ (see Table~\ref{table:bayes}). This conclusion is robust to the choice of priors: The peak of the posterior column density probability from MODEL C (which has $b_{\rm{OVI}}$ fixed to the X--ray best--fit value), lies within $13<$log~$N_\mathrm{OVI}$~(cm$^{-2})=<14$, while $99.7$~\% of the distribution falls below log~$N_\mathrm{OVI}$~(cm$^{-2})=13.98$. 
In case of MODEL D, which applies the X--ray measurement based $b_{\rm{OVI}}$ prior, the corresponding $3\sigma$ upper limit for log~$N_\mathrm{OVI}$~(cm$^{-2})$ increases to $15.00$, as broader line profiles are allowed. 
The Bayesian analysis thus indicates that whereas the two--phase scenario is consistent with the measurement data, the one--phase alternative (explaining both the \ion{O}{VI} and \ion{O}{VII} detections) appears unlikely.





\subsubsection{Broad HI Lyman-alpha}

In addition, the CIE model predicts a \ion{H}{I} Ly$\alpha$ line
with $b_\mathrm{HI}=194_{-19}^{+46}$~km$\,$s$^{-1}$, and log$N_\mathrm{HI}($cm$^{-2})=12.9\pm0.2\, \times \,$log$ Z_\sun /Z$. 
The signal of this line would overlap with the blend of the two detected Ly$\alpha$ lines at $z=0.04512$ and $z=0.04567$ (Table~\ref{table:FUV}),
effectively preventing the spectral modeling of this line.

Therefore, we follow an alternative method that consists of calculating the statistical significance level, $N_\sigma$, expected for the X--ray predicted Lyman alpha line present in STIS data by applying the following equation \citep[see][for details]{hellsten98}:
\begin{equation}\label{eq:ew}
    \mathrm{EW_{HI}}=\frac{N_\sigma\sqrt{N_\mathrm{pix}}\Delta \lambda_\mathrm{pix}}{\mathrm{S/N}\,(1+z_\mathrm{HI})}.
\end{equation}
Here, $N_\mathrm{pix}$ is the number of detector pixels used in the EW detection, and $\Delta \lambda_\mathrm{pix}$ denotes the pixel widths in the spectral wavelength units. Throughout our calculations, we adopt $N_\mathrm{pix}$ value that corresponds the FWTM (full width at 1/10 maximum) of a Gaussian ($b_\mathrm{HI}=200$~km$\,$s$^{-1}$) line profile convolved with the instrumental LSF (see Sect.~\ref{fuv_an}).

We calculate 3 and 5$\sigma$ EW limits from the continuum signal-to-noise ratio using Eq.~\ref{eq:ew}, and convert them into column density limits, assuming the line will fall in the linear part of curve of growth. Since the X--ray predicted, broad Ly$\alpha$ line will coincide with the absorption by narrower Ly$\alpha$ features from cooler gas phases, the data cannot be used to directly calculate the upper limits. Thus, we first model out the cooler phase HI contribution (using the best--fit, two--component model presented in Table~\ref{table:FUV}) and use then the standard deviation of the residual continuum fluxes to calculate the desired upper limits.

\begin{figure}[ht]
\begin{center}
\includegraphics[width=3.6in]{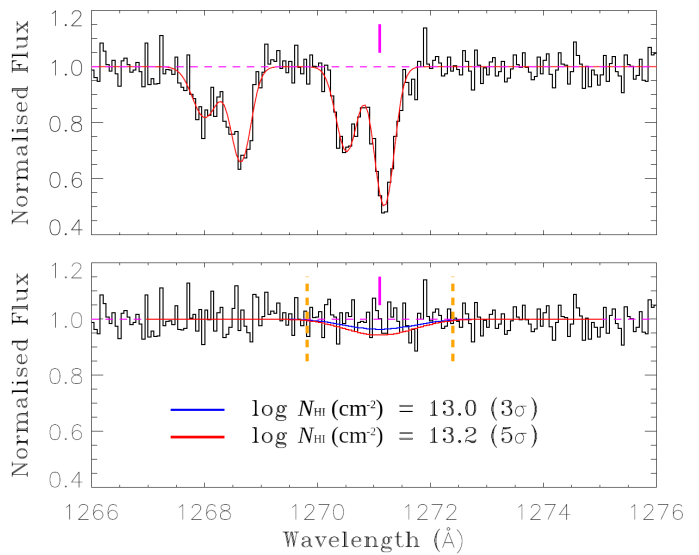}
\end{center}
\caption{Upper limits on column density of a broad \ion{H}{I} Ly$\alpha$. Top panel: Model of the (narrow) \ion{H}{I} Ly$\alpha$ blend used in the upper limit determination. Bottom panel: Normalized residual spectrum after model removal. Red and blue curves show the broad ($b_\mathrm{HI}=200$~km$\,$s$^{-1}$) line profiles at limiting column densities as indicated. Purple vertical lines mark the expected centroid wavelength of the broad line, as based on the X--ray prediction.
}
\label{fig:broadHI}
\end{figure}

The analysis yields formal $3\sigma$ upper limit $\log N_\mathrm{HI}$~(cm$^{-2})\lesssim13.0$, and a $5\sigma$ limit $\log N_\mathrm{HI}$~(cm$^{-2})\lesssim13.2$ for the broad HI (see Fig.~\ref{fig:broadHI}). Comparing these limits to the best-fit CIE model predicted $N_\mathrm{HI}$, we calculate the corresponding lower limits for the hot phase metallicity, for which we get $Z\gtrsim0.8\times Z_\sun$ and $Z\gtrsim 0.5\times Z_\sun$, respectively.

The obtained metallicity limits appear high for intergalactic gas, for which one would generally expect $Z\sim0.01-0.1\times Z_\sun$, or $Z\lesssim0.4\times Z_\sun$ in the regions close to galactic halos \citep[e.g.,][]{martizzi19}.
While we acknowledge that our approach includes several sources of uncertainties, such as those involved in the $z=0.04512$, $z=0.04567$ Ly$\alpha$ blend model, and the uncertainties related to the X--ray model predicted BLA line profile (e.g., in $z_\ion{H}{I}$, $N_\mathrm{HI}$, $b_\mathrm{HI}$ and the CIE model O/H), the BLA analysis seems to exclude the possibility of much lower metallicities (e.g., that of a typically assumed WHIM metallicity, $Z\approx 0.1\times Z_\sun$).

It is worth noticing, however, that our limits on metallicity are limited to the measurements of hot, \ion{O}{VII} absorbing phase only. It is very well possible that the obtained metallicity limit is not representative 
outside this phase.
Indeed, as shown in Fig~13 in \cite[]{wijers20}, 
the \texttt{EAGLE} simulation predicts highly ionized oxygen absorbers' ion-weighted metallicities to remain atypically high (i.e., close to $\approx Z_\sun$) up to at least a few $r_{200}$'s distances from galaxies. While our metallicity limit on the \ion{O}{VII} absorbing phase agrees with this prediction, it may also reflect a more general situation when it comes to the observational studies of high column density metal line absorbers; the instrumental sensitivity limits drive a selection effect that favors detections of high metallicity gas.


\begin{figure}[ht]
\begin{center}
\includegraphics[width=3.6in]{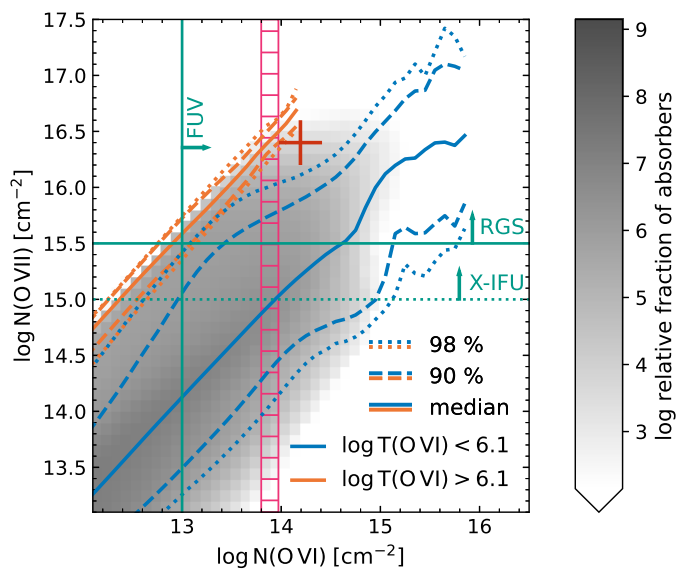}
\end{center}
\caption{Ion column density distribution of co--located \ion{O}{VI} and \ion{O}{VII} in \texttt{EAGLE} (gray histogram in the background). The orange contours show the percentiles of $N_\mathrm{OVII}$ that are counterparts to \ion{O}{VI} absorbers with mean ion-weighted \ion{O}{VI} temperature $>10^{6.1}$~K (hot phase) as a function of the total (i.e., for all phases) \ion{O}{VI} column density. The blue contours show the same for $T_\mathrm{OVI}<10^{6.1}$~K (warm phase). The pink bars indicate the FUV measured warm \ion{O}{VI} column density $1\sigma$ uncertainty range, whereas the red point marks the total ion columns of the two--phase model (i.e., hot and warm \ion{O}{VI}, hot \ion{O}{VII}). The green lines mark the approximate detection thresholds for \ion{O}{VI} and \ion{O}{VII} in FUV and X--ray bands. Detection thresholds for two representative X--ray instruments are shown: the RGS camera on--board \emph{XMM--Newton}, and the Athena X-IFU, currently under development.
}
\label{fig:coldens}
\end{figure}

\section{Comparison to the EAGLE simulation}\label{disc}

The \ion{O}{VI} and \ion{O}{VII} column densities inferred from the two-phase model (Sect.~\ref{two-phase})  
are now compared to predictions from the \texttt{EAGLE} cosmological, hydrodynamical simulation \citep[][]{schaye15,crain15,mcalpine16} for co-spatial \ion{O}{VI} and \ion{O}{VII} gas in the cosmic web.  

In Fig.~\ref{fig:coldens}, we show the simulated distributions of \ion{O}{VI} and \ion{O}{VII} ion column densities through $z=0.1$ absorption systems  in \texttt{EAGLE}
\citep[for details on generating the simulated data, we refer to][]{wijers19}. 
In the context of this work, and with regards to X--ray follow-up studies of FUV detected WHIM absorbers in general, we are particularly interested in examining the \texttt{EAGLE} predictions for typical \ion{O}{VII} ion columns co-located with hot/warm \ion{O}{VI} absorbers.
To study this, we divide the \texttt{EAGLE} ($N_\mathrm{OVI}$, $N_\mathrm{OVII}$) -distribution into `warm' and `hot' sub--distributions by applying a temperature cut based on the ion-weighted \ion{O}{VI} temperature, $T_\mathrm{OVI}^{i}$, through the \texttt{EAGLE} absorbers. The temperature cut was performed at $T_\mathrm{OVI}^{i}=10^{6.1}$~K, according to the derived $3\sigma$ upper limit on the FUV-detected, warm phase temperature (Sect.~\ref{cie}). We note that the ion-weighted temperatures in \texttt{EAGLE} are averaged for all phases together, and not limited to cooler, detectable gas.

\begin{figure*}[t]
\begin{center}
\begin{minipage}{0.52\textwidth}
\includegraphics[width=\textwidth]{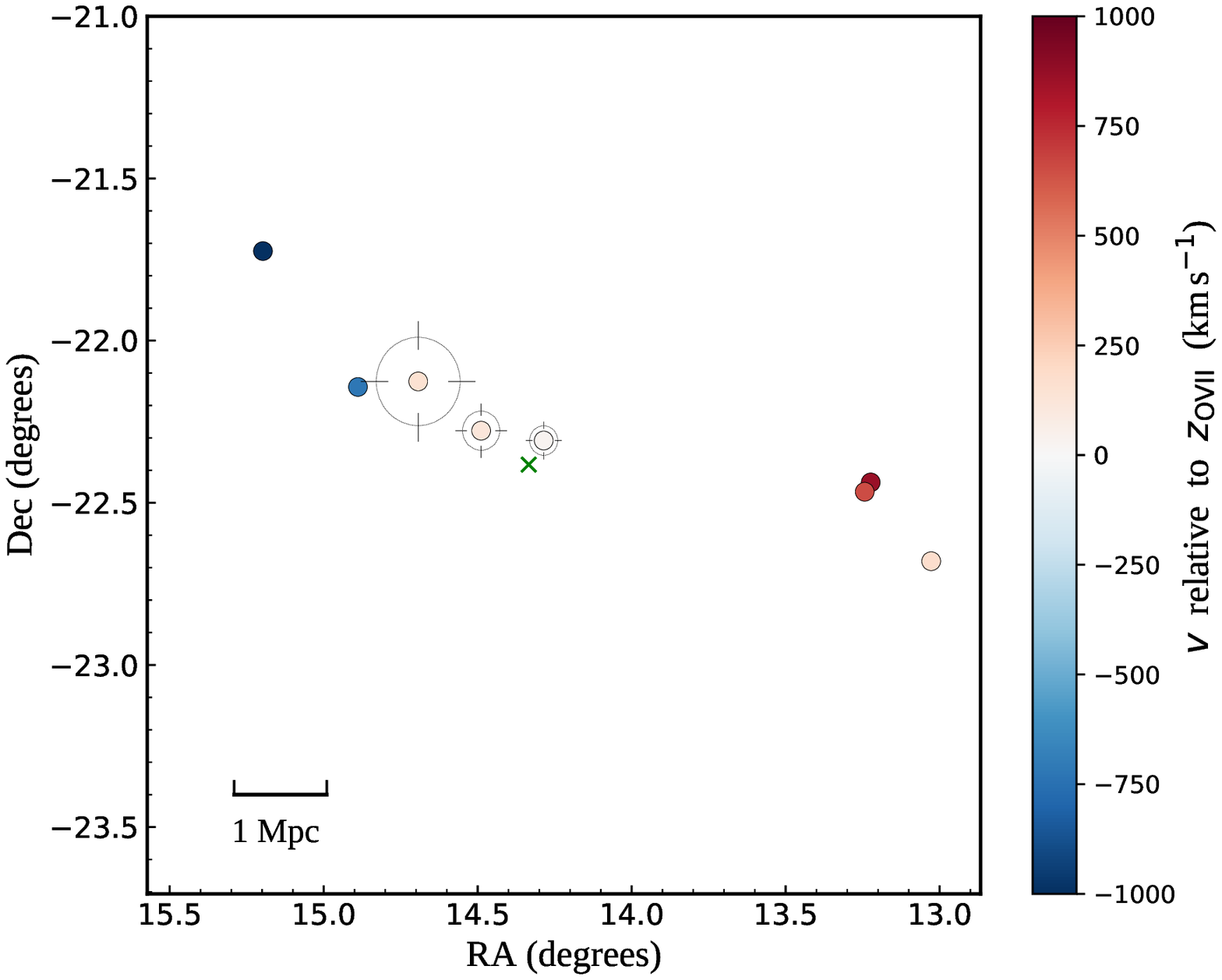}
\end{minipage}
\begin{minipage}{0.47\textwidth}
\includegraphics[width=\textwidth]{{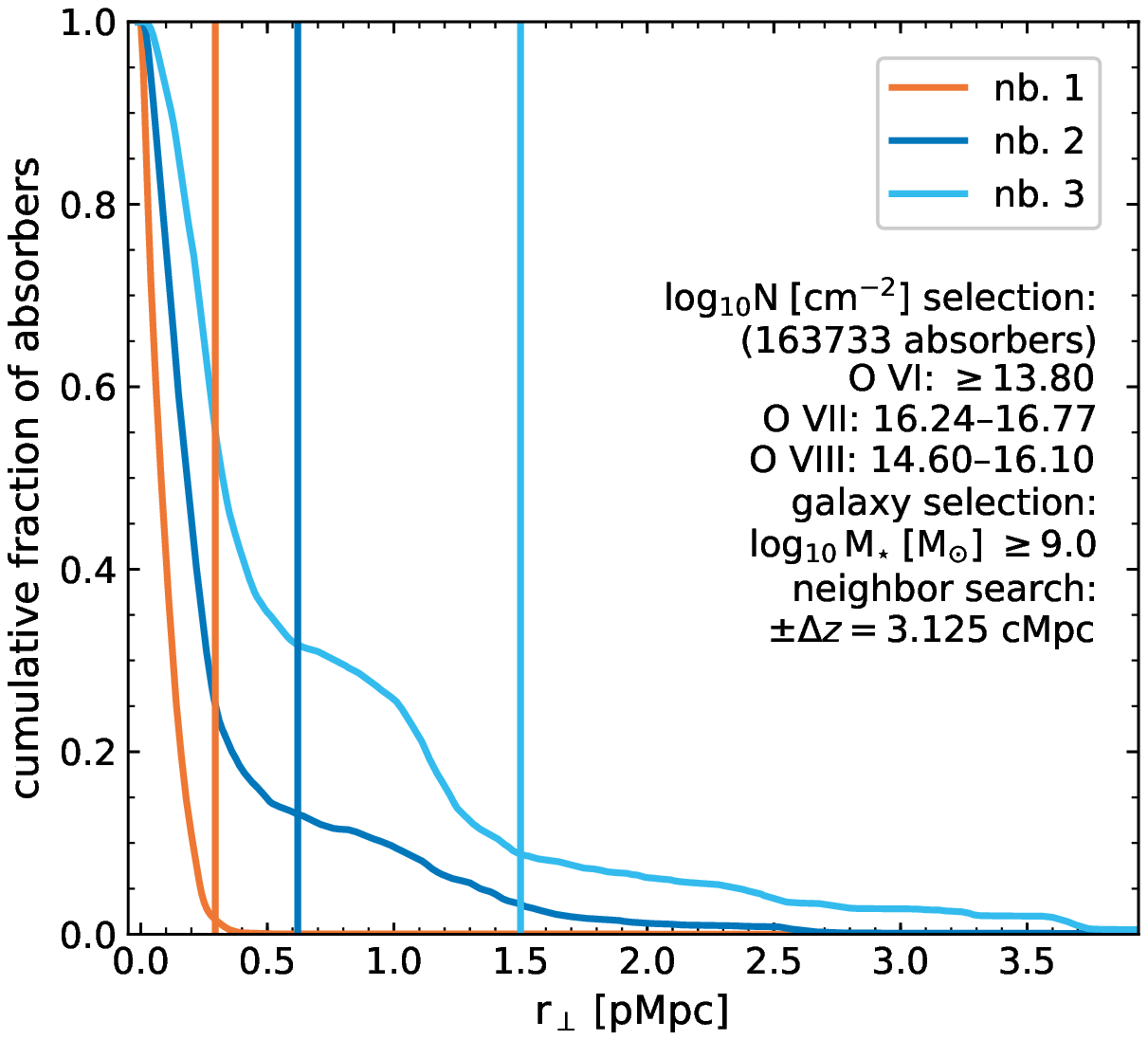}}
\end{minipage}

\end{center}
\caption{Left: Galactic environment near the examined absorber. The green cross is the Ton~S~180 sightline and the filled circles mark all the detected galaxies within $|\delta z|<0.005$ (corresponding to radial depth $|\delta r|< 2$~cMpc) of $z_\mathrm{OVII}$. Color coding corresponds to relative radial velocities with respect to $z_\mathrm{OVII}$ as calculated from the spectroscopic redshifts. The circles around the three galaxies with the smallest impact parameters to the sightline show the predicted $R_{200}$s along with the expected $68$~\% variation limits (marked with the bars). The proper distance scale at $z_\mathrm{OVII}$ is shown.
Right: The fraction of \texttt{EAGLE} absorbers like those observed for which the galaxy with $M_*\gtrsim 10^{9.0} M_\sun$ and $|\delta r_\mathrm{||}|<3.125$~cMpc that is closest to the sightline has an impact parameter greater that the value plotted along the x-axis (nb.1: orange curve). The blue and light blue curve show the same for the galaxies with the second and third smallest impact parameters, respectively. The vertical lines show the impact parameters of the three nearest galaxies found in this work (see left panel). We note that the apparently large sample of \texttt{EAGLE} absorbers represents only a small fraction of the total volume ($1 \times 10^{-5}$). 
}
\label{fig:environment}
\end{figure*}

As illustrated in Fig.~\ref{fig:coldens}, the distributions for warm and hot \ion{O}{VI} appear to be very different.
The warm phase distribution (blue contours) is
notably broader, in this parameter space, than that of the hotter phase (orange contours).
The two distributions do not overlap within their respective 90 percentile limits, with the hotter absorbers producing systematically higher $N_\mathrm{OVII}$ at a given $N_\mathrm{OVI}$ throughout the examined parameter space. 
At the FUV--measured $N_\mathrm{OVI}$ level (pink bars), only a small fraction of warm \ion{O}{VI} absorbers have a detectable \ion{O}{VII} counterpart, whereas this fraction would increase towards higher \ion{O}{VI} column densities.
Even though the warm phase produces relatively low \ion{O}{VII} ion columns, the median of the warm phase ($N_\mathrm{OVI}$, $N_\mathrm{OVII}$) -distribution has $N_\mathrm{OVII}/N_\mathrm{OVI}\approx10^1$, and there are no absorption systems with
$N_\mathrm{OVII}/N_\mathrm{OVI}<10^{-1}$, showing that \ion{O}{VII} is more abundant than \ion{O}{VI} in the \texttt{EAGLE} absorption systems, at least in the examined parameter range. 
We note that a similar conclusion about the \ion{O}{VI}, \ion{O}{VII} relative abundances can be drawn from the \texttt{magneticum} simulations of the oxygen ion mass distributions in the WHIM, presented in \cite{khabibullin19}.

The two-phase model data point agrees with the hot, $T_\mathrm{OVI}^i>10^{6.1}$~K absorption system oxygen distribution, but not with the warm one. This indicates that the CIE assumption adopted for the \ion{O}{VII} absorbing phase is generally supported by \texttt{EAGLE}. 
In \texttt{EAGLE}, column densities as high as those found for \ion{O}{VIII} and especially \ion{O}{VII} mostly occur in the CGM of relatively massive galaxies ($\gtrsim 10^{12} \mathrm{M}_\sun$). 
Indeed, the high-end \ion{O}{VII} column densities are expected to be found inside galactic halos, and in line with this, \texttt{EAGLE} predicts a steep drop in \ion{O}{VII} ion column densities when the sightline impact parameter exceeds $\rho \approx R_{200}$, due to decreasing $n_\mathrm{H}$ and metallicity \citep{wijers20}.

To investigate whether halo absorption can explain the measurements for the Ton~S~180 absorber, we study the galactic environment surrounding the sightline near $z_\mathrm{OVII}$.
In the left panel of Fig.~\ref{fig:environment}, we plot all the galaxies with known redshifts around the absorber within a volume corresponding to $9\times9$~pMpc$^2$ transverse and $4$~cMpc in depth. These galaxies appear to form a thread--like structure in the 2D--projection, stretching at least a few Mpc in length, while passing the sightline at a close distance. We note that such a spatial distribution of the nearby galaxies could imply that the sightline intercepts a filament of a cosmic web at $z\approx0.0456$, but we cannot confirm this scenario without additional data which is not available at this time.

To study the origin of the oxygen absorption, we estimate the virial radii of the galactic halos with the smallest distances to the sightline (and $z$ within the uncertainty limits of $z_\mathrm{OVII}$). To estimate the stellar masses, $M_*$, for the three closest galaxies, we compared the available $B$-, $R$- and $K$-band photometric data \citep[adopted from][]{1996MNRAS.278.1025L,2006AJ....131.1163S,prochaska11} to the color--stellar mass distributions of the galaxies with matching absolute magnitudes in the COSMOS2015 galaxy sample \citep{2016ApJS..224...24L}. We obtain the most likely stellar masses and their 68th percentile limits, for which we get $10^{9.5\pm0.5}$, $10^{10.1\pm0.5}$ and $10^{10.95\pm0.2}M_\sun$, as listed in order of increasing impact parameter to the sightline. 
These stellar masses are converted into the halo mass estimations by adopting the best--fit model to the median observed $M_*-M_h$ -relationship in the $z=0.1$ Universe, as presented in \citet{2019MNRAS.488.3143B}.
The derived $M_h$ estimates are used to calculate the corresponding $R_{200}$'s for each of the closest galaxies. The results of the galactic environment analysis are listed in Table~\ref{table:gal_env}, and the obtained $R_{200}$'s illustrated with circles in the left panel of Fig.~\ref{fig:environment}. 
We find that CGM absorption by the detected nearby galaxies cannot explain the X--ray absorption. 



\begin{table*}[ht]\label{table:gal_env}
\caption{Galactic environment around the Ton~S~180 oxygen absorber.}
\begin{center}
\begin{tabular}{lccccc}
\hline
\hline
Galaxy ID                  & $z$ & log$\,M_*$  &  log$\,M_h$ &  $R_{200}$ (kpc)  & $\rho/R_{200}$  \\ 
\hline
2MASX J00570854-2218292    & $0.04562$ & $9.5\pm0.5$ & $11.5\pm0.3$ & $160_{-30}^{+40}$ & $1.8\pm0.4$ \\ 
PWC2011 J005757.3-221640   & $0.04596$ & $10.1\pm0.5$ & $11.8_{-0.3}^{+0.5}$ & $210_{-45}^{+80}$ & $2.9\pm0.8$ \\ 
MCG-04-03-036              & $0.04608$ & $10.95\pm0.20$ & $12.9\pm0.4$ & $480_{-135}^{+170}$ & $3.1_{-0.8}^{+1.2}$ \\ 
\hline
\end{tabular}
\tablefoot{Observed and derived properties for the three galaxies with the smallest impact parameters, $\rho$, to the sightline near the $z_\mathrm{OVII}=0.0455\pm0.0005$ absorber. The masses are given in units of Solar mass.}
\end{center}
\end{table*}

Therefore, and given the apparent galaxy overdensity in the vicinity of the $z\approx0.0456$ absorber, we examine the occurrence of high \ion{O}{VII-VIII} ion column absorbers in different galactic environments in \texttt{EAGLE}. In practice, we inspect the correlation between strong \ion{O}{VII-VIII} absorbers and their nearest neighboring $M_*\gtrsim 10^{9.0} M_\sun$ galaxies within $6.25$~cMpc thick slices through the \texttt{EAGLE}. In the analysis, the galactic environment is quantified in terms of the 2D (perpendicular) distances of the nearest galaxies to the sightlines. 
We look for galaxies close to absorbers in \texttt{EAGLE}, rather than the other way around, for two (related) reasons. Firstly, it mimics the selection effects in the observations, where we looked for \ion{O}{VI} absorption, then X--ray absorption where we found it, and finally galaxies where we found both. Secondly, absorbers as strong as those we found here, especially for \ion{O}{VII}, are rare in \texttt{EAGLE}, and they are rare around galaxies of all masses. However, weaker absorbers might not have been detected, so we do not want to base our comparison too much on how rare such simulated absorption systems may be. Quantifying environments of any strong absorbers that do exist in the simulation avoids this. 

In the right panel of Fig.~\ref{fig:environment} we consider only the subsample of the \texttt{EAGLE} absorbers fulfilling each of the measurement constraints discussed in this work: the FUV measured lower limit on $N_\mathrm{OVI}$ and the `slab' constraints for \ion{O}{VII-VIII} ion columns measured at the same redshift. The orange curve labeled as "nb. 1" shows the fraction of absorbers from this sample for which the nearest galaxy (with $M_*\gtrsim 10^{9.0} M_\sun$ and $|\delta r_\mathrm{||}|<3.125$~cMpc) has an impact parameter greater than the value plotted along the $x$-axis. The blue curves show the same but for the impact parameter of the second closest (nb. 2), and the third closest (nb. 3), galaxy.
The analysis reveals that even though the high ion column densities are most often located in close proximity to the galaxies (indicating CGM absorption), the X--ray detectable metal columns span a wide range of environments.
The figure also indicates, for instance, that up to $\approx10$~\% of the considered \texttt{EAGLE} absorbers occur in environments similar to those we found surrounding the Ton~S~180 absorber (orange and blue vertical lines). 
Overall, our analysis suggests that the strong \ion{O}{VII} X--ray absorber in Ton~S~180 sightline originates in the transition region between the CGM and the diffuse WHIM.

Finally, we look back to Fig.~\ref{fig:coldens} to interpret it from an observational point of view. We infer that the \texttt{EAGLE} spatial correlation between warm \ion{O}{VI} and \ion{O}{VII} (and \ion{O}{VIII}, not shown here) 
is likely a manifestation of the complex multiphase structures of the intergalactic absorbers in \texttt{EAGLE}.
The narrow \ion{O}{VI} lines 
can therefore be used to trace hot WHIM gas associated with the same multi--temperature structure, 
whereas the broad
\ion{O}{VI} lines could reveal the direct counterparts for hot and collisionally ionized absorbers.
However, the vast majority of the current FUV detections of intergalactic \ion{O}{VI} absorbers have $b_\mathrm{OVI}<100$~km$\,$s$^{-1}$, with the peak of distribution at $b_\mathrm{OVI}\approx30$~km$\,$s$^{-1}$ \citep[e.g.,][]{sembach01, danforth08, tripp08, tilton12, danforth16}, whereas
in this work we measured a non-thermal line broadening component $b_\mathrm{nt}=97_{-49}^{+77}$~km$\,$s$^{-1}$ for the hot phase. 
We note that if such high $b_\mathrm{nt}$ values are 
typical for the X--ray detectable absorbers, then the hot CIE \ion{O}{VI} counterparts may often be undetectable in the FUV, 
due to their broad and shallow line profiles. As discussed in Sect.~{\ref{two-phase}}, the detectability of \ion{O}{VII} lines decreases steeply towards smaller $b$-values, and therefore the prospects of simultaneous detection of both \ion{O}{VI} and \ion{O}{VII} from a common WHIM phase may be unlikely with the current observational instruments. As a result, the method utilizing \ion{O}{VI} detections to trace X--ray absorbing WHIM seems generally more useful for finding absorption systems with complex thermal structures. 

Whereas the number of X--ray detectable WHIM absorbers will likely remain low within the performance limits of current instruments,
future X--ray instruments with higher sensitivity, such as the Athena X--IFU \citep{barret18}, 
should reveal a much larger sample of \ion{O}{VII} WHIM absorbers (see Fig.~\ref{fig:coldens}), with the potential to resolve the missing baryons problem.
With statistical data on hot absorbers' ion column densities and their line-of-sight gas velocity dispersions, combined with the parallel analysis of the associated FUV absorbers, information on the heating mechanisms and evolution of the WHIM could ultimately be obtained.


\section{Conclusions}

In this paper we have presented 
the detection of a WHIM X--ray absorber towards the NLS1 
galaxy Ton~S~180, discovered by following up on the FUV (\ion{O}{VI}) detection at redshift  $z_\mathrm{OVI}=0.04579$. Our main conclusions are the following: \begin{enumerate}[i]

\item The measurements of
\ion{O}{VII} (and upper limit on \ion{O}{VIII}) absorption line fluxes are consistent with a collisional ionization equilibrium state, characterized by $T_\mathrm{CIE}=1.7\pm0.2\times10^6$~K and $N_\mathrm{H}\times Z_{\sun}/Z_\mathrm{abs}=5.8_{-2.2}^{+3.0}\times10^{19}$~cm$^{-2}$. The X--ray redshift of the absorber, $z_\mathrm{OVII}=0.0455\pm0.0005$, is consistent with two FUV \ion{O}{VI} absorbers locating at $z=0.04557$ and $z=0.045979$. The latter one of these FUV absorbers is characterized by \ion{O}{VI} ion column density matching the CIE model prediction. We link the oxygen absorption to a filamentary concentration of galaxies observed at a consistent redshift (left panel of Fig.~\ref{fig:environment}).

\item The FUV \ion{O}{VI} absorption at $z=0.04579\pm0.00001$ is described by log$N_\mathrm{OVI}$~(cm$^{-2})=13.68\pm0.10$ and a $3\sigma$ upper limit for Doppler parameter $b_\mathrm{OVI}<38.1$~km$\,$s$^{-1}$. 
This \ion{O}{VI} absorber cannot be
the counterpart to the \ion{O}{VII}--absorbing phase, as is indicated by the \ion{O}{VI} line width constraints on the ion temperature (yielding a $3\sigma$ upper limit $T_\mathrm{OVI}<1.4\times10^6$~K), and by the inconsistency of $b_\mathrm{OVI}$ with that of the \ion{O}{VII} producing phase, for which we measure $b_\mathrm{OVII}=106_\mathrm{-42}^{+73}$~km$\,$s$^{-1}$. The hot phase \ion{O}{VI} line is not detectable within the S/N of the current FUV data, because of the shallow and broad line profile.

\item Our interpretation is that the X--ray and FUV signals are due to a multiphase 
absorber located at/close to an CGM/IGM transition zone of a complex, large scale galactic structure. 
The FUV measured \ion{O}{VI} column density through this absorption system is $\sim10^{-3}$ times $N_\mathrm{OVII}$, indicating that 
the information on the absorbing oxygen (metal) content is imprinted mainly in the X--ray band. 
The $N_\mathrm{OVI}-N_\mathrm{OVII}$ distribution in the \texttt{EAGLE} cosmological, hydrodynamical simulation suggests that this is the case for all X--ray detectable intergalactic \ion{O}{VII} absorbers, 
and also for a large fraction of absorption systems currently identified only by FUV \ion{O}{VI} absorption. Our \texttt{EAGLE} analysis also implies that the X--ray detectable WHIM absorbers are mainly to be found in the vicinity of structures with enhanced galaxy densities.

\end{enumerate} 

The results presented in this work highlight the important role 
of line width measurements in the interpretation of physical conditions in the WHIM. 
Line width also affects detectability: The broad \ion{O}{VI} associated with typical \ion{O}{VII} absorbers requires very high quality FUV spectra, while the narrow \ion{O}{VII} associated with typical \ion{O}{VI} absorbers is hard to detect with X--ray telescopes because the line saturation reduces the equivalent width.
We conclude that the line widths will be the key to understanding the physical nature and more accurately estimating the baryon content of highly ionized WHIM absorbers, when higher sensitivity X--ray instruments enable the examination of a large sample of X--ray WHIM absorbers.

\begin{acknowledgements}
JA and FA acknowledge the financial support received under ESA Athena grant 3880057 and thank J. Huovelin for supporting the X-IFU science case. SM is supported by the Experienced Researchers fellowship from the Alexander von Humboldt-Stiftung, Germany. We thank E. Churazov for discussions regarding the interpretation of the \texttt{magneticum} WHIM simulation data.
\end{acknowledgements}

\bibliographystyle{aa} 
\bibliography{references}

\end{document}